\documentclass[journal]{IEEEtran}

\ifCLASSINFOpdf \usepackage[pdftex]{graphicx} \else \fi 

\usepackage[cmex10]{amsmath}
\usepackage{amsthm}
\usepackage{amssymb}
\usepackage{verbatim}
\usepackage{mathdots}
\usepackage{mathtools}

\usepackage{psfrag}

\usepackage{cite}

\usepackage{mathrsfs}
\usepackage[utf8]{inputenc}
\usepackage[T1]{fontenc}

\usepackage{enumitem}
\usepackage{dsfont}

\usepackage{caption}
\usepackage[position=b]{subcaption}
\DeclareCaptionLabelSeparator{periodspace}{.\quad}
\newcounter{muni}
\newenvironment{remunerate}
               {\begin{list}{{\upshape 
               \arabic{muni}.}}{\usecounter{muni}
                \setlength{\leftmargin}{0pt}
                \setlength{\itemindent}{25pt}}}{\end{list}}

\makeatletter
\newcommand{\labitem}[2]{%
\def\@itemlabel{#1}
\item
\def\@currentlabel{#1}\label{#2}}
\makeatother

\begin{document}

\title{Why RLC realizations of certain impedances need many more energy storage elements than expected}
\author{Timothy H. Hughes \thanks{This research was conducted while the author was the Henslow research fellow at Fitzwilliam College, University of Cambridge, U.K., supported by the Cambridge Philosophical Society, http://www.cambridgephilosophicalsociety.org.} 
\thanks{Timothy H.\ Hughes is with the Department of Engineering, University of Cambridge, Trumpington Street, Cambridge, UK, CB2 1PZ, e-mail: thh22@cam.ac.uk.}\thanks{\copyright \hspace{0.1cm} 2017 IEEE. This is the accepted version of the manuscript: Hughes, T.H.: Why RLC realizations of certain impedances need many more energy storage elements than expected. IEEE Trans.\ Autom.\ Control, 62(9), 4333-4346 (2017).}}


\newtheorem{theorem}{Theorem}
\newtheorem{lemma}[theorem]{Lemma}
\newtheorem{proposition}[theorem]{Proposition}
\newtheorem{corollary}[theorem]{Corollary}
\newtheorem{definition}[theorem]{Definition}
\newtheorem{remark}[theorem]{Remark}
\providecommand{\abs}[1]{\left\lvert#1\right\rvert}

\maketitle

\begin{abstract}
It is a significant and longstanding puzzle that the resistor, inductor, capacitor (RLC) networks obtained by the established RLC realization procedures appear highly non-minimal from the perspective of linear systems theory. Specifically, each of these networks contains significantly more energy storage elements than the McMillan degree of its impedance, and possesses a non-minimal state-space representation whose states correspond to the inductor currents and capacitor voltages. Despite this apparent non-minimality, there have been no improved algorithms since the 1950s, with the concurrent discovery by Reza, Pantell, Fialkow and Gerst of a class of networks (the RPFG networks), which are a slight simplification of the Bott-Duffin networks. Each RPFG network contains more than twice as many energy storage elements as the McMillan degree of its impedance, yet it has never been established if all of these energy storage elements are necessary. In this paper, we present some newly discovered alternatives to the RPFG networks. We then prove that the RPFG networks, and these newly discovered networks, contain the least possible number of energy storage elements for realizing certain positive-real functions. In other words, \emph{all} RLC networks which realize \emph{certain} impedances contain more than twice the expected number (McMillan degree) of energy storage elements.
\end{abstract}

\begin{IEEEkeywords}
Passive system, positive-real, minimal realization, network synthesis, electric circuit, mechanical control, inerter.
\end{IEEEkeywords}

\section{Introduction}
\label{sec:intro}
\IEEEPARstart{M}{odern} systems theory has its roots in electrical circuit analysis and synthesis \cite[p.\ 78]{JWBAOIS}. The notions of realizability, minimality, and the relationship between the internal and external properties of systems, all feature in several classical papers on electrical circuit synthesis, e.g., \emph{Foster's reactance theorem} \cite{Foster_24}. The connection between passivity and positive-real (PR) functions originated in the thesis of Otto Brune on electrical circuit synthesis \cite{Brune}, where it was established that the impedance of any passive network is necessarily PR. These concepts continue to play a central role in modern systems theory. Nevertheless, many significant results in electrical circuit synthesis continue to perplex systems theorists. In particular, there remain several open questions on the synthesis of PR impedances with networks comprising resistors, inductors, and capacitors (RLC networks). Some of the more puzzling questions concern \emph{minimality} (in terms of the numbers and types of elements required) \cite{kalman2010, HugJSmOp}; \emph{controllability} \cite{camwb, JWDDS}; and \emph{observability} \cite{JW_HVDS, JWDDS}. Notably, an RLC network can contain more energy storage elements than the McMillan degree of its impedance, and possess a non-minimal state-space representation whose states correspond to the inductor currents and capacitor voltages. Indeed, this is the case for the famous Bott-Duffin networks \cite{BD}, and their simplifications \cite{Reza_54, PS, Fialkow_Gerst}. The purpose of this paper is to demonstrate the necessity of this apparent non-minimality in the RLC realization of certain PR functions.

In \cite{HugSmSP}, it was established that the Bott-Duffin networks contain the least possible number of energy storage elements for realizing certain PR functions (the \emph{biquadratic minimum functions}) using \emph{series-parallel} networks. However, it is possible to realize an arbitrary given PR function with RLC networks which are not series-parallel and contain fewer energy storage elements than the Bott-Duffin networks. This is demonstrated by the networks discovered by Reza, Pantell, Fialkow and Gerst \cite{Reza_54, PS, Fialkow_Gerst} (hereafter referred to as the \emph{RPFG networks}), which achieve a slight improvement on the Bott-Duffin networks. In this paper, we first present some newly discovered alternatives to the RPFG networks. We then prove that, among the \emph{entire} class of RLC networks, the RPFG networks and our newly discovered alternatives contain the least possible number of energy storage elements for realizing almost all biquadratic minimum functions. This is despite the number of energy storage elements in these networks being more than twice the McMillan degree of the corresponding network's impedance. 

Secondary to the motivation outlined above, the topic of this paper is also relevant to mechanical control following the recent invention of the \emph{inerter} \cite{mcs02}. Using the completed electrical-mechanical analogy (see Appendix \ref{sec:spmc}), any given RLC network has a corresponding damper-spring-inerter network whose transfer function from force to velocity is equal to the impedance of the corresponding RLC network. Such damper-spring-inerter networks have applications in vibration absorption systems \cite{mcs02}, e.g., vehicle suspension \cite{chen_14}; train suspension \cite{fucheng_10, fucheng_12, jiang_vsd_12}; motorcycle steering compensators \cite{Limebeer_steering2, Limebeer_Steering}; and building suspension \cite{fucheng_07}.

The structure of this paper, and the key contributions, are as follows. Section \ref{sec:ssrrlc} discusses state-space descriptions of RLC network behaviors. In Section \ref{sec:ibdp}, we present the RPFG networks (Fig.\ \ref{fig:ps}), and our newly discovered alternatives (Fig.\ \ref{fig:nabd}). Each network in Section \ref{sec:ibdp} contains significantly more energy storage elements than the McMillan degree of its impedance. In Section \ref{sec:mbdi}, we state our main results concerning the necessity of this apparent non-minimality for the realization of biquadratic minimum functions (Theorems \ref{thm:bmfr3re}--\ref{thm:bmfr5re}). Section \ref{sec:rmfrlc} investigates the realization of general (not necessarily biquadratic) minimum functions with RLC networks. The main results are then proved in Section \ref{sec:rbmfrlc}.

Relevant background information is included in three appendices. Appendices \ref{sec:nc} and \ref{sec:tnab} contain technical information on RLC network classification and the graph theoretic analysis of RLC networks. The reader who wishes to follow the proofs in Sections \ref{sec:rmfrlc} and \ref{sec:rbmfrlc} in detail is advised to read these appendices before those sections. Finally, in Appendix \ref{sec:spmc}, we outline the electrical-mechanical analogy and its relevance to this paper.

Our notation is as follows. We let $\mathbb{R}$ (resp., $\mathbb{C}$) denote the real (resp., complex) numbers. For $z \in \mathbb{C}$, we denote the real (resp., imaginary) part by $\Re{(z)}$ (resp., $\Im{(z)}$), and the complex conjugate of $z$ by $z^{*}$. $\mathbb{R}[s]$ (resp.,  $\mathbb{R}(s)$) denotes the polynomials (resp., rational functions) in the indeterminate $s$ with real coefficients. With $\mathbb{F}$ denoting one of $\mathbb{R}$, $\mathbb{C}$, $\mathbb{R}[s]$, or $\mathbb{R}(s)$, then $\mathbb{F}^{m \times n}$ and $\mathbb{F}^{n}$ denote matrices and vectors of the respective dimensions whose entries are all from $\mathbb{F}$. We let $\text{diag}\!\begin{pmatrix}M_{1}& \hspace{-0.15cm} \cdots & \hspace{-0.15cm} M_{n}\end{pmatrix}$ denote the block diagonal matrix with $M_{1}, \ldots , M_{n}$ on the main diagonal; and $\text{col}\!\begin{pmatrix}M_{1}& \hspace{-0.15cm} \cdots & \hspace{-0.15cm} M_{n}\end{pmatrix} := \begin{bmatrix}M_{1}^{T}& \cdots& M_{n}^{T}\end{bmatrix}^{T}$. For $G \in \mathbb{R}(s)$, we say $G$ is PR if (i) $G$ is analytic in the open right half plane; and (ii) $\Re{(G(\lambda))} \geq 0$ for $\Re{(\lambda)} > 0$. Equivalently, condition (ii) can be replaced with (iii) $\Re{(G(j\omega))} \geq 0$ for all $\omega \in \mathbb{R}$ (except at poles of $G$), and the poles of $G$ on $j\mathbb{R} \cup \infty$ are simple and have real positive residues. $G$ is called lossless if it is PR and $\Re{(G(j\omega))} = 0$ for all $\omega \in \mathbb{R} \cup \infty$. When $\hat{p}, \hat{q} \in \mathbb{R}[s]$ are coprime, and $G = \hat{p}/\hat{q}$, then the McMillan degree of $G$ is the maximum of the degrees of $\hat{p}$ and $\hat{q}$, and is equal to the number of states in a minimal (controllable and observable) state-space realization for $G$.

\section{State-space representations of RLC network behaviors}
\label{sec:ssrrlc}
The famous Bott-Duffin networks \cite{BD}, and their simplifications \cite{Reza_54, PS, Fialkow_Gerst} (the \emph{RPFG networks}), prove that any given PR function can be realized by an RLC network. However, the number of energy storage elements in these networks is considerably greater than the McMillan degree of their impedance. In contrast, there are many RLC networks which possess the same number of energy storage elements as the McMillan degree of their impedance. For example, any \emph{regular function} of McMillan degree two (biquadratic) can be realized by an RLC network containing two energy storage elements \cite{JiangSmith11}. Indeed, in the analysis of electrical networks, it is not unusual to assume that the behavior of the network has a minimal state-space realization whose states correspond to the inductor currents and capacitor voltages (see \cite[Section III]{BelOc} for a nice description of this and other commonly held assumptions and their implications). In this section, we provide examples of networks which violate this condition. One conclusion of this paper is that this assumption is violated for \emph{all} RLC realizations of \emph{certain} PR functions (the biquadratic minimum functions). In fact, we will answer the question \emph{what is the minimum possible number of energy storage elements required for realizing a biquadratic minimum function}, and we find that in most cases the RPFG networks actually contain the least possible number of energy storage elements.

\begin{figure}[!t]
 \begin{center}
\leavevmode
\includegraphics[page=1,width=0.28\hsize]{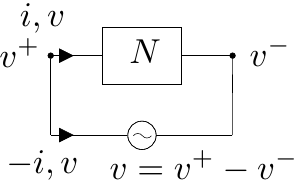}
\end{center}
\caption{RLC network with source. The driving-point current and voltage are denoted by $i$ and $v$, respectively.}
\label{fig:open}
\end{figure}

\begin{figure}[!t]
\scriptsize
  \begin{center}
    \leavevmode
\includegraphics[page=2,width=0.88\hsize]{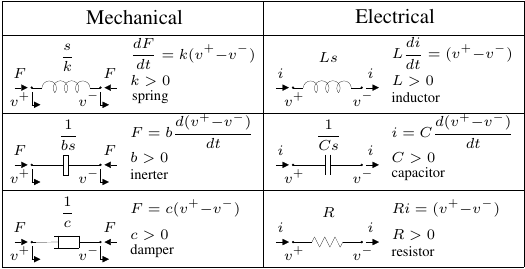}
\end{center}
\caption{Passive electrical and mechanical elements.}
\label{fig:ema}
\end{figure}

We consider RLC (one-port) networks. These networks possess a pair of driving-point terminals across which a source can be attached as in Fig.\ \ref{fig:open}. The network comprises an interconnection of resistors, inductors, and capacitors, which have the properties shown in Fig.\ \ref{fig:ema} (we only allow strictly positive values for $R$, $L$, and $C$). Note that this figure also indicates the similarities between these elements and three mechanical components: springs, dampers, and inerters (this will be discussed in Appendix \ref{sec:spmc}).

The driving-point current $i$ and voltage $v$ are constrained by the network to satisfy a linear differential equation of the form $p(\tfrac{d}{dt})\mathbf{i} = q(\tfrac{d}{dt})\mathbf{v}$ for some $p, q \in \mathbb{R}[s]$. Providing $q \not\equiv 0$, then the \emph{impedance} $Z$ of the network is defined as $Z := p/q$. The existence of $Z$ is guaranteed if there is at least one path of elements between the driving-point terminals of the network \cite{HugNa}, in which case $Z$ is PR, and the number of energy storage elements in the network is greater than or equal to the McMillan degree of $Z$ \cite{HugSmAI}.

As emphasised in \cite{camwb, JW_HVDS, HugNa}, certain RLC networks contain more energy storage elements than the McMillan degree of their impedance. There are two main ways in which this can happen: (i) it may not be possible to arbitrarily assign initial values to the currents through the inductors and the voltages across the capacitors; and (ii) the set of inductor currents and capacitor voltages can be either uncontrollable or unobservable from the driving-point terminals. 

\begin{figure}[!t]
\centering 
\subcaptionbox{Network with linearly dependent capacitor voltages. \label{fig:cap_circ}}{
\centering 
\scriptsize
\includegraphics[page=8,width=0.45\hsize]{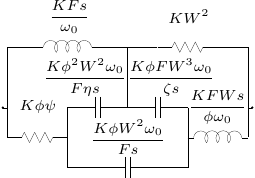}
}
\hspace{0.1cm}
\subcaptionbox{Network whose behavior is not stabilizable. \label{fig:cap_cs}}{
\centering 
\hspace*{-0.3cm}
\scriptsize
\includegraphics[page=6,width=0.48\hsize]{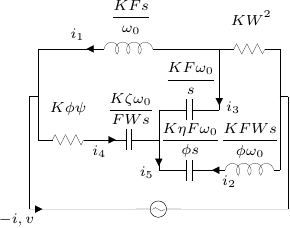}
}
\caption{Two network realizations of the function $H(s)$ in Lemma \ref{lem:bmfp}; $\phi := 1-W$, $\psi := 1+W$, $\eta := 2W-1$, $\zeta := W^{2}\phi^{2}-F^{2}\eta$, $K, \omega_{0} > 0$ $0<F < W(1-W)/\sqrt{2W-1}$, $1/2 < W < 1$. The two networks are related through a star-delta transformation involving the three capacitors.} \label{fig:probn} 
\end{figure}

Case (i) is illustrated by the network in Fig.\ \ref{fig:cap_circ}. The three capacitors in this network form a circuit, so the sum of the voltages across the capacitors must sum to zero by Kirchhoff's voltage law. It follows that the behavior of this network does not possess a state-space representation $\tfrac{d\mathbf{x}}{dt} = A\mathbf{x} + Bi$, $v = C\mathbf{x} + Di$ whose states correspond to the inductor currents and capacitor voltages. To see this, note that for any given (real) $\mathbf{x}(0)$ and (locally integrable) $i$, then $\mathbf{x}(t) = e^{At}\mathbf{x}(0) + \int_{0}^{t}e^{A(t-\tau)}Bi(\tau)d\tau$ and $v(t) = C\mathbf{x}(t) + Di(t)$ ($t \geq 0$) is a solution to the state-space equations. If the states correspond to the inductor currents and capacitor voltages, then this implies that there are trajectories of the network for which the initial voltages across the capacitors do not sum to zero: a contradiction. Note, however, that it is possible to describe the behavior of this network using a linear differential algebraic equation \cite{HugNa}.

Case (ii) is illustrated by the network in Fig.\ \ref{fig:cap_cs}. The behavior of this network possesses the state-space representation: $\tfrac{d\mathbf{x}}{dt} = A\mathbf{x} + Bi$, $v = C\mathbf{x} + Di$; with $\mathbf{x} := \begin{bmatrix}i_{1}& i_{2}& v_{3}& v_{4}& v_{5}\end{bmatrix}^{T}$,
\begin{align*}
&A {:=} \!\left[\!\begin{smallmatrix}-\tfrac{\omega_{0}\phi \psi}{F}& 0& \tfrac{\omega_{0}}{FK} & -\tfrac{\omega_{0}}{FK}& 0 \\ 0& -\tfrac{\omega_{0}W\phi}{F}& -\tfrac{\omega_{0}\phi}{FWK}& 0& -\tfrac{\omega_{0}\phi}{FWK} \\ -\omega_{0}FK& -\omega_{0}FK& 0& 0& 0\\ \tfrac{\omega_{0}K\zeta}{FW}& 0& 0& 0& 0 \\ 0& -\tfrac{\omega_{0}KF\eta}{\phi}& 0& 0 & 0 \end{smallmatrix}\!\right]\!, \hspace{0.05cm} B {:=} \!\left[\!\begin{smallmatrix}-\tfrac{\omega_{0}\phi \psi}{F} \\ -\tfrac{\omega_{0}W\phi}{F} \\ -\omega_{0}FK \\ \tfrac{\omega_{0}K \zeta}{FW}\\ 0\end{smallmatrix}\!\right]\! \\
&C := \left[\!\begin{smallmatrix}K\phi\psi & \hspace{0.1cm} KW^{2} & \hspace{0.1cm}  -1& \hspace{0.1cm}  1& \hspace{0.1cm}  0\end{smallmatrix}\!\right]\!, \text{ and } D:= K,
\end{align*} 
and where $\phi, \psi, \eta$, and $\zeta$ are as defined in the caption of Fig.\ \ref{fig:probn}. Now, let 
\begin{equation*}
\mathbf{\tilde{x}} = \begin{bmatrix}-\phi W^{2}& \phi^{2}W& -KF^{2}\eta & K\zeta & KF^{2}\eta \end{bmatrix}^{T}.
\end{equation*}
Then $A\mathbf{\tilde{x}} = -\tfrac{W \phi}{F}\mathbf{\tilde{x}}$ and $C\mathbf{\tilde{x}} = 0$, so this state-space model is not observable.  Similarly, with $\mathbf{\hat{x}} = \begin{bmatrix}0 & 0& \eta \zeta & F^{2}W\eta & -\phi \zeta \end{bmatrix}^{T}$, then $\mathbf{\hat{x}}^{T}A = 0$ and $\mathbf{\hat{x}}^{T}B = 0$, so this state-space model is not controllable. In fact, this state-space model is not stabilizable owing to an uncontrollable mode at the origin (as emphasised in \cite{HugNa}, this violates an assumption which is implicit in the a-c steady-state analysis of RLC networks adopted in \cite{Smf}).

In this paper, we investigate the necessity of the apparent non-minimality of RLC networks such as those in Fig.\ \ref{fig:probn} for the RLC realization of certain PR functions. In fact, we will show that each network in Fig.\ \ref{fig:probn} contains the least possible number of energy storage elements for realizing its impedance.

\section{RLC Network Realization Procedures}
\label{sec:ibdp}

The famous Bott Duffin procedure \cite{BD} provided the first algorithm for realizing a general positive-real function as the impedance of an RLC network. Slight simplifications of the Bott Duffin networks were discovered by Reza, Pantell, Fialkow and Gerst in the 1950s \cite{Reza_54, PS, Fialkow_Gerst} (the \emph{RPFG networks}). In this section, we present some newly discovered alternatives to these networks. It will follow from the results in Section \ref{sec:mbdi} that the RPFG networks, and these new alternatives, contain the least possible number of energy storage elements for realizing almost all \emph{biquadratic minimum functions} (defined below). 
\begin{definition}
$H(s) \in \mathbb{R}(s)$ is called a minimum function (with \emph{minimum frequency} $\omega_{0}$) if $H$ is PR, not identically zero, has no poles or zeros on $j \mathbb{R} \cup \infty$, and satisfies $\Re \left(H(j\omega_{0})\right) = 0$ for at least one $\omega_{0} > 0$ (which implies $\Im \left(H(j\omega_{0})\right) \neq 0$). It is called \emph{biquadratic} if its McMillan degree is two.
\end{definition}

\begin{figure*}[!t]
\centering 
\begin{subfigure}{1.0\columnwidth}
\centering 
\scriptsize
\includegraphics[page=12,width=0.80\columnwidth]{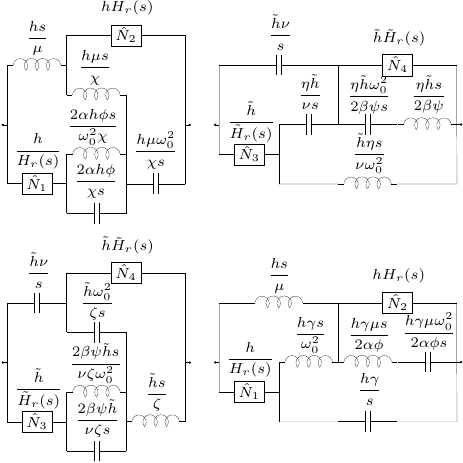}
\caption{The RPFG networks for realizing a minimum function.}
\label{fig:ps}
\end{subfigure}
\begin{subfigure}{1.0\columnwidth}
\centering 
\scriptsize
\includegraphics[page=13,width=0.80\columnwidth]{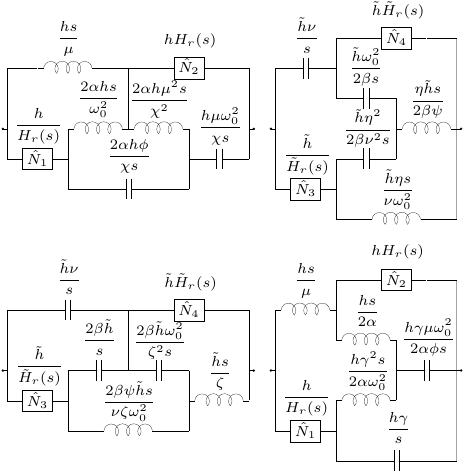}
\caption{Newly discovered alternative networks.}
\label{fig:nabd}
\end{subfigure}
\caption{Illustration of a single step in the realization of a minimum function $H(s)$ with $H(j\omega_{0}) = \omega_{0}Xj$. Here, $h = H(\mu)$, $\chi = \omega_{0}^{2} + 2\alpha \mu$, $\gamma = \mu + 2\alpha$, $\phi = \chi + \mu^{2}$, $\tilde{h} = H(\nu)$, $\eta = \omega_{0}^{2} + 2\beta \nu$, $\zeta = \nu + 2\beta$, $\psi = \eta + \nu^{2}$, and $H_{r}(s)$, $\mu$, $\alpha$, $\tilde{H}_{r}(s)$, $\nu$, $\beta$ are defined in the proof of Theorem \ref{thm:rmf}.} \label{fig:ex3} 
\end{figure*}

Both the Bott-Duffin procedure and the RPFG simplification are inductive, with two stages at each inductive step: 1) the problem of realizing an arbitrary given PR function $G(s)$ is converted into the problem of realizing a minimum function, derived from $G(s)$, whose McMillan degree is no greater than that of $G(s)$; 2) the problem of realizing an arbitrary given minimum function $H(s)$ is converted into the problem of realizing two PR functions, derived from $H(s)$, whose McMillan degrees are at least two fewer than that of $H(s)$. Stage 1 (described above) is achieved by the \emph{Foster preamble}, as discussed in \cite[Section II]{HugSmSP}. The contribution of \cite{Reza_54, PS, Fialkow_Gerst} was the discovery of the networks in Fig.\ \ref{fig:ps}, which pertain to stage 2. In Fig.\ \ref{fig:nabd}, we present the newly discovered alternatives to these networks. As shown in \cite[Section 3.1]{HugTh}, the networks in Fig.\ \ref{fig:ex3} can be derived from the Bott-Duffin networks by a sequence of network transformations (see also \cite{Storer_54} for an alternative sequence of transformations relating the networks in Fig.\ \ref{fig:ps} to the Bott-Duffin networks). As a consequence of the following theorem, the networks in Fig.\ \ref{fig:ex3} can be used in stage 2 of the procedure described above.
\begin{theorem}
\label{thm:rmf}
Let $H(s)$ be a minimum function with $H(j\omega_{0}) = \omega_{0}Xj$ and $X > 0$ (resp., $X < 0$). Then $H(s)$ is realized as the impedance of the networks on the top left and bottom right (resp., top right and bottom left) of Figs.\ \ref{fig:ps} and \ref{fig:nabd} for some $\alpha, \mu > 0$ (resp., $\beta, \nu > 0$), and some PR function $H_{r}(s)$ (resp., $\tilde{H}_{r}(s)$) whose McMillan degree is at least two fewer than that of $H(s)$.
\end{theorem}

\begin{IEEEproof}
Consider first the case $X > 0$. Then, as described in \cite[Section II]{HugSmSP}, there exists a $\mu > 0$ such that $X = H(\mu)/ \mu$, and there exists an $\alpha > 0$ and a PR function $H_{r}(s)$ such that $(\mu H(\mu) - sH(s))/(\mu H(s) - s H(\mu)) = 1/H_{r}(s) + 2\alpha s/(s^{2}+\omega_{0}^{2})$. Here, the McMillan degree of $H_{r}(s)$ is at least two fewer than that of $H(s)$. It follows that 
\begin{equation*}
\label{eq:rtei}
H(s) = H(\mu)\frac{s^{3} + H_{r}(s)(2\alpha + \mu)s^{2} + \omega_{0}^{2}s + H_{r}(s) \mu \omega_{0}^{2}}{H_{r}(s)s^{3} + \mu s^{2} + H_{r}(s) (2\alpha \mu + \omega_{0}^{2})s + \mu \omega_{0}^{2}}.
\end{equation*}
Direct calculation verifies that this is the impedance of the networks on the top left and bottom right of Figs.\ \ref{fig:ps} and \ref{fig:nabd}.

If, instead, $X < 0$, then there exists a $\nu > 0$ such that $-\omega_{0}^{2} X = H(\nu) \nu$, and there exists a $\beta > 0$ and a PR function $\tilde{H}_{r}(s)$ such that $(\nu H(s) - s H(\nu))/(\nu H(\nu) - sH(s)) = 1/\tilde{H}_{r}(s) + 2\beta s/(s^{2}+\omega_{0}^{2})$ \cite[Section II]{HugSmSP}. In this case, the McMillan degree of $\tilde{H}_{r}(s)$ is at least two fewer than that of $H(s)$, and we obtain
\begin{equation*}
H(s) = H(\nu)\frac{\tilde{H}_{r}(s)s^{3}+\nu s^{2} + \tilde{H}_{r}(s)(\omega_{0}^{2} + 2\beta \nu)s + \nu \omega_{0}^{2}}{s^{3} + \tilde{H}_{r}(s)(2\beta + \nu) s^{2} + \omega_{0}^{2}s + \tilde{H}_{r}(s)\nu \omega_{0}^{2}}.
\end{equation*}
By direct calculation, this is the impedance of the networks on the top right and bottom left of Figs.\ \ref{fig:ps} and \ref{fig:nabd}.
\end{IEEEproof}

\section{RLC realizations and minimality}
\label{sec:mbdi}

Each network in Fig.\ \ref{fig:ex3} contains many more energy storage elements than expected for the realization of its impedance. However, we will prove that these networks contain the least possible number of energy storage elements (five) and the least possible number of resistors (two) for realizing almost all biquadratic minimum functions. Our main results are stated in Theorems \ref{thm:bmfr3re}--\ref{thm:bmfr5re}, which adopt the parametrisation of a biquadratic minimum function described in the following lemma: 
\begin{lemma}
\label{lem:bmfp}
Let $H(s)$ be a biquadratic minimum function. Then $H(s)$ takes the form
\begin{equation*}
K\frac{s^{2} + \frac{\omega_{0}\left(1-W\right)F}{W}s+\omega_{0}^{2}W}{s^{2} + \frac{\omega_{0}\left(1-W\right)}{F}s + \frac{\omega_{0}^{2}}{W}},
\end{equation*}
for some $K, \omega_{0} > 0$ and for some $W, F$ which satisfy either (i) $0<W<1$ and $F>0$, or (ii) $W>1$ and $F<0$. Here $K, \omega_{0}, F$, and $W$ are uniquely determined by $H(s)$, with $K = H(\infty)$, $KW^{2} = H(0)$, and $KFj = H(j\omega_{0})$.
\end{lemma}
\begin{IEEEproof}
This is immediate from \cite[Theorem 8]{Smf}. Relative to the terminology of \cite[equations (4), (8)]{Smf}, we have made the substitutions $R = K$, $k = W$, and $X_{0} = FK$ (this enables a more concise presentation of the main results).
\end{IEEEproof}

\begin{figure}[!h]
\scriptsize
\begin{center}
\includegraphics[page=14,width=0.72\hsize]{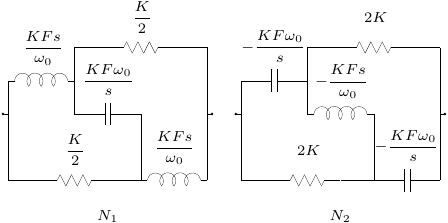}
\end{center}
\caption{Networks $N_{1}$ and $N_{2}$. The impedances of both $N_{1}$ and $N_{2}$ have the form indicated in Lemma \ref{lem:bmfp}; and $F$ satisfies condition (a) (resp., (b)) of Theorem \ref{thm:bmfr3re} in network $N_{1}$ (resp., $N_{2}$).}
\label{fig:q28}
\end{figure}

\begin{theorem}
\label{thm:bmfr3re}
Let $N$ be an RLC network whose impedance $H(s)$ is a biquadratic minimum function, as in Lemma \ref{lem:bmfp}. Then $N$ contains at least three energy storage elements and at least two resistors. If, in addition, $N$ contains exactly three energy storage elements, then either (a) $W = \tfrac{1}{2}$ and $F > 0$, or (b) $W = 2$ and $F < 0$. In particular, $H(s)$ is the impedance of $N_{1}$ (resp., $N_{2}$) in case (a) (resp., (b)) (see Fig.\ \ref{fig:q28}).
\end{theorem}

\begin{figure}[!t]
\scriptsize
\begin{center}
\includegraphics[page=15,width=0.75\hsize]{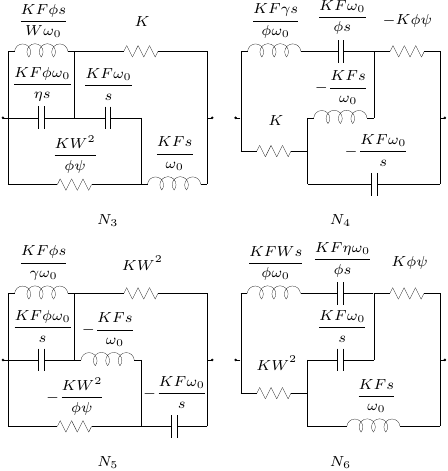}
\end{center}
\caption{Networks $N_{3}$, $N_{4}$, $N_{5}$, and $N_{6}$. Here, $\phi = 1-W$, $\psi = 1+W$, $\eta = 2W-1$, and $\gamma = 2-W$.  The impedances of all four networks have the form indicated in Lemma \ref{lem:bmfp}, and $F$ and $W$ satisfy condition \ref{nl:4rebmfc1} (resp., \ref{nl:4rebmfc2}, \ref{nl:4rebmfc3}, \ref{nl:4rebmfc4}) of Theorem \ref{thm:bmfr4re} in network $N_{3}$ (resp., $N_{4}$, $N_{5}$, $N_{6}$).}
\label{fig:q29}
\end{figure}

The proof of Theorem \ref{thm:bmfr3re} comes at the end of Section \ref{sec:rbmfrlc}, as does the proof of the following theorem: 
\begin{theorem}
\label{thm:bmfr4re}
Let $N$ be an RLC network whose impedance $H(s)$ is a biquadratic minimum function, as in Lemma \ref{lem:bmfp}. If $N$ contains four or fewer energy storage elements, then either condition (a) or (b) in Theorem \ref{thm:bmfr3re} is satisfied, or one of the following four conditions must hold:
\begin{enumerate}[label=(\alph*)]
\setcounter{enumi}{2}
\item $1/2 < W < 1$ and $F = W\sqrt{2W-1}/(1-W)$, \label{nl:4rebmfc1}
\item $1 < W < 2$ and $F = (1-W)\sqrt{W/(2-W)}$, \label{nl:4rebmfc2}
\item $1 < W < 2$ and $F = \sqrt{W^3(2-W)}/(1-W)$, \label{nl:4rebmfc3}
\item $1/2 < W < 1$ and $F = W(1-W)/\sqrt{2W-1}$. \label{nl:4rebmfc4}
\end{enumerate}
In particular, $H(s)$ is the impedance of $N_{3}$ (resp., $N_{4}$, $N_{5}$, $N_{6}$) in case \ref{nl:4rebmfc1} (resp., \ref{nl:4rebmfc2}, \ref{nl:4rebmfc3}, \ref{nl:4rebmfc4}) (see Fig.\ \ref{fig:q29})
\end{theorem}

We then obtain the following theorem (also proved at the end of Section \ref{sec:rbmfrlc}):
\begin{theorem}
\label{thm:bmfr5re}
Let $N$ be an RLC network whose impedance $H(s)$ is a biquadratic minimum function, as in Lemma \ref{lem:bmfp}. Then the following conditions all hold. 
\begin{enumerate}[label=\arabic*., ref=\arabic*, leftmargin=0.4cm] 
\item If none of the conditions (a)--(f) in Theorems \ref{thm:bmfr3re} and \ref{thm:bmfr4re} are satisfied, then $N$ contains at least five energy storage elements and at least two resistors.\label{nl:5recon1}
\item If $F > 0$ and $0 < W < 1$, then $H(s)$ is the impedance of the networks on the top left and bottom right of Figs.\ \ref{fig:ps} and \ref{fig:nabd}, where $X = FK/\omega_{0}$, $\mu = W\omega_{0}/F$, $H(\mu) = KW$, $\alpha = (F^{2} + W^{2})(1-W)\omega_{0}/(2W^{2}F)$, $H_{r}(s) = W$, and $\hat{N}_{1}$ and $\hat{N}_{2}$ are both resistors.\label{nl:5recon2}
\item If $F < 0$ and $W > 1$, then $H(s)$ is the impedance of the networks on the top right and bottom left of Figs.\ \ref{fig:ps} and \ref{fig:nabd}, where $X = FK/\omega_{0}$, $\nu = -F\omega_{0}/W$, $H(\nu) = KW$, $\beta = (F^{2} + W^{2})(1-W)\omega_{0}/(2WF)$, $\tilde{H}_{r}(s) = 1/W$, and $\hat{N}_{3}$ and $\hat{N}_{4}$ are both resistors.\label{nl:5recon3}
\end{enumerate}
\end{theorem}

We also note that the impedance of each network in Fig.\ \ref{fig:probn} has the form indicated in Lemma \ref{lem:bmfp}, with $K, \omega_{0} > 0$, $0<F < W(1-W)/\sqrt{2W-1}$, and $\tfrac{1}{2} < W < 1$. It then follows from Theorem \ref{thm:bmfr5re} that each network in Fig.\ \ref{fig:probn} contains the least possible number of energy storage elements and the least possible number of resistors for the realization of its impedance. This is despite the fact that the capacitor voltages are linearly dependent in the network in Fig.\ \ref{fig:cap_circ}, and the behavior of the network in Fig.\ \ref{fig:cap_cs} is not stabilizable.

To show Theorems \ref{thm:bmfr3re}--\ref{thm:bmfr5re}, we first determine those minimum functions (not necessarily biquadratic) which are realized by RLC networks containing four or fewer energy storage elements (this results in Theorems \ref{thm:mfr34re} and \ref{thm:mfr4re} in Section \ref{sec:rmfrlc}). The proofs of Theorems \ref{thm:bmfr3re}--\ref{thm:bmfr5re} (at the end of Section \ref{sec:rbmfrlc}) then amount to determining which biquadratic minimum functions are realized by the networks described in Theorems \ref{thm:mfr34re} and \ref{thm:mfr4re}. 

We note an important distinction between Theorems \ref{thm:bmfr3re}--\ref{thm:bmfr5re} and the results in the paper \cite{Smf}. Specifically, Theorems \ref{thm:bmfr3re}--\ref{thm:bmfr5re} establish the minimum possible number of energy storage elements in RLC networks realizing certain PR impedances, whereas \cite{Smf} considers the minimum possible number of elements. As discussed in Section \ref{sec:ssrrlc}, the number of energy storage elements is more relevant from a linear systems theory perspective. Moreover, Theorems \ref{thm:bmfr3re} and \ref{thm:bmfr4re} cover a class of infinitely many networks (as there is no restriction on the number of resistors), whereas the results in \cite{Smf} only cover a class of finitely many networks (those containing seven or fewer elements). 

In fact, there are two notable errors in the main results of \cite{Smf}. First, \cite[Theorem 12]{Smf} was shown to be incorrect by Foster, who stated `perhaps such a census (of biquadratic minimum functions realized by RLC networks containing seven or fewer elements) should still be included among those problems not completely solved as yet' \cite{Foster_63c}. Second, in \cite[p.\ 349]{Smf}, it is claimed that the RPFG  networks are `the only general seven-element realizations of the biquadratic minimum function'. But this is disproved by the networks in Fig.\ \ref{fig:nabd} of this paper. More specifically, the seven-element RPFG networks (see Fig.\ \ref{fig:ps} and Theorem \ref{thm:bmfr5re}) realize the set of all biquadratic minimum functions, i.e., the set of all $H(s)$ of the form indicated in Lemma \ref{lem:bmfp}. If $0 < W < 1$ and $F > 0$, then $H(s)$ is the impedance of the networks on the top left and bottom right of Fig.\ \ref{fig:ps} \emph{and also Fig.\ \ref{fig:nabd}} (see condition \ref{nl:5recon2} of Theorem \ref{thm:bmfr5re}); and if $W > 1$ and $F < 0$, then $H(s)$ is the impedance of the RPFG networks on the top right and bottom left of Fig.\ \ref{fig:ps} \emph{and also Fig.\ \ref{fig:nabd}} (see condition \ref{nl:5recon3} of Theorem \ref{thm:bmfr5re}). We note that the networks in Figs.\ \ref{fig:probn}, \ref{fig:q28}, and \ref{fig:q29} (and other networks in \cite{HugTh}) realize impedances $H(s)$ of the form indicated in Lemma \ref{lem:bmfp} for some but not all possible values of $W$ and $F$, so they are not `general realizations of the biquadratic minimum function'.

The realization of biquadratic minimum functions with RLC networks containing seven or fewer elements was reconsidered in \cite{HugTh}, and the networks in Figs.\ \ref{fig:probn} and \ref{fig:nabd} were discovered. There, it is shown that the networks in Fig.\ \ref{fig:ex3} are the only seven-element realizations of \emph{certain} biquadratic minimum functions, e.g., $(s^{2} + \tfrac{1}{2}s + \tfrac{2}{3})/(s^{2} + \tfrac{1}{3}s + \tfrac{3}{2})$, which has the form indicated in Lemma \ref{lem:bmfp} with $K = \omega_{0} = F = 1$ and $W = \tfrac{2}{3}$.

\section{RLC realizations of minimum functions}
\label{sec:rmfrlc}
In this section, we investigate the realization of general (not necessarily biquadratic) minimum functions with RLC networks containing limited numbers of energy storage elements. We state the main results in Theorems \ref{thm:mfr34re} and \ref{thm:mfr4re}, which adopt the network classification scheme described in Appendix \ref{sec:nc}.

\begin{figure}[!h]
\scriptsize
\begin{center}
\includegraphics[page=16,width=0.6\hsize]{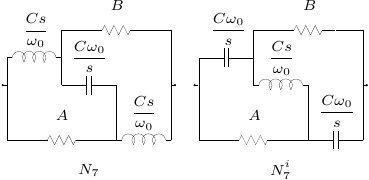}
\end{center}
\caption{Quartet $\mathcal{Q}_{7}$: $A, B, C > 0$.}
\label{fig:mfrn1}
\end{figure}

\begin{theorem}
\label{thm:mfr34re}
Let $N$ be an RLC network whose impedance $H(s)$ is a minimum function. Then $N$ contains at least three energy storage elements. Moreover, if $N$ contains exactly three energy storage elements, then $H(s)$ is the impedance of a network from $\mathcal{Q}_{7}$ (see Fig.\ \ref{fig:mfrn1}).
\end{theorem}

\begin{figure}[!b]
\scriptsize
\begin{center}
\includegraphics[page=17,width=0.65\hsize]{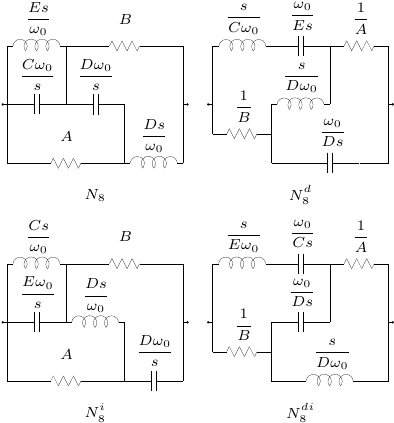}
\end{center}
\caption{Quartet $\mathcal{Q}_{8}$: $A, B, C, D > 0; E \coloneqq \frac{CD}{C+D}$.}
\label{fig:mfrn2}
\end{figure}

\begin{figure}[!b]
\scriptsize
\begin{center}
\includegraphics[page=18,width=0.65\hsize]{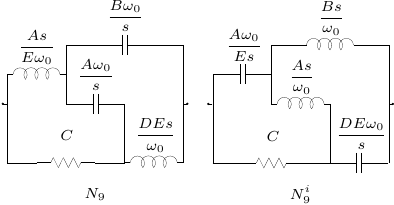}
\end{center}
\caption{Quartet $\mathcal{Q}_{9}$: $A, B, C, D > 0; E \coloneqq \frac{A+B}{B+D}$.}
\label{fig:mfrn3}
\end{figure}

\begin{figure}[!t]
\scriptsize
\begin{center}
\includegraphics[page=19,width=0.3\hsize]{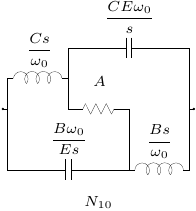}
\end{center}
\caption{Quartet $\mathcal{Q}_{10}$: $A, B, C, (B-D)(C-D) > 0, B \neq C; E \coloneqq \frac{B-D}{C-D}$.}
\label{fig:mfrn4}
\end{figure}

\begin{figure}[!t]
\scriptsize
\begin{center}
\includegraphics[page=20,width=0.68\hsize]{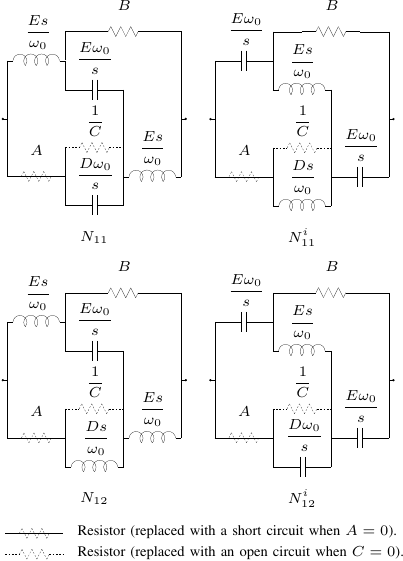}
\end{center}
\caption{Network classes $\mathcal{N}_{11}$, $\mathcal{N}_{11}^{i}$, $\mathcal{N}_{12}$, and $\mathcal{N}_{12}^{i}$: $A, B, C, D, E > 0$; $\mathcal{N}_{11a}$, $\mathcal{N}_{11a}^{i}$, $\mathcal{N}_{12a}$, and $\mathcal{N}_{12a}^{i}$: $A = 0$, $B, C, D, E > 0$; $\mathcal{N}_{11b}$, $\mathcal{N}_{11b}^{i}$, $\mathcal{N}_{12b}$, and $\mathcal{N}_{12b}^{i}$: $C = 0$, $A, B, D, E > 0$} 
\label{fig:mfrn5}
\end{figure}

\begin{theorem}
\label{thm:mfr4re}
Let $N$ be an RLC network whose impedance $H(s)$ is a minimum function. If $N$ contains four or fewer energy storage elements, then $H(s)$ is the impedance of a network from one of the classes defined in Figs.\ \ref{fig:mfrn1} to \ref{fig:mfrn5}
\end{theorem}

\begin{figure}[!t]
\scriptsize
\begin{center}
\includegraphics[page=21,width=0.35\hsize]{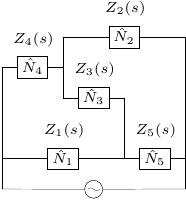}
\end{center}
\caption{Network described in Lemma \ref{lem:mfffre}.} 
\label{fig:ns1}
\end{figure}

The proofs of Theorems \ref{thm:mfr34re} and \ref{thm:mfr4re} come at the end of this section, and rely on Lemmas \ref{lem:pe}--\ref{lem:mfffre}. The structure of the argument is as follows. We let $N$ be a network whose impedance satisfies the properties required of a minimum function, and we consider a \emph{sinusoidal trajectory} of $N$ at the frequency $\omega_{0}$ (i.e., the driving-point and internal element current and voltages all vary sinusoidally at this frequency). From equation (\ref{eq:s0rpe}), the energy supplied to $N$ over a single period of the sinusoidal trajectory is equal to the energy dissipated in $N$ over the same interval. But the properties of a minimum function imply that the energy supplied to $N$ over a single period is zero, so there is no energy dissipated in $N$. In particular, there are no currents flowing through the resistors in $N$. More generally, there are subnetworks within $N$ which are \emph{blocked}, i.e., they have no current flowing through them, and their vertices are all at the same potential. Also, the only unblocked elements are energy storage elements. We then find that if $N$ contains four or fewer energy storage elements, then $N$ must comprise the one-ports $\hat{N}_{1}$ to $\hat{N}_{5}$ connected as in Fig.\ \ref{fig:ns1} (e.g., $N_{11}$ in Fig.\ \ref{fig:mfrn5} has this form, where $\hat{N}_{1}$ is the one-port corresponding to the parallel connection of a resistor and a capacitor in series with another resistor). Further conditions (mainly relating to the absence of poles and zeros at the origin and infinity) then result in Theorems \ref{thm:mfr34re} and \ref{thm:mfr4re}.

To formalise this argument, we use the hierarchical graph theory based approach to the analysis of RLC networks outlined in Appendix \ref{sec:tnab} (see also \cite{HugTh, HugNa}). The reader who wishes to follow the detailed proofs in this section is advised to read that appendix now. It contains numbered notes (\ref{nl:gt0-}, \ref{nl:gt0}, etc) which will be referred to in the proofs. Formal definitions for the terminology in this section can also be found in that appendix (e.g., subnetwork, one-port, sinusoidal trajectory, driving-point trajectory, phasor current and voltage). We emphasise here the distinction between a \emph{subnetwork} and a \emph{one-port} (which is a special type of subnetwork). For example, the two capacitors in network $N_{11}$ in Fig.\ \ref{fig:mfrn5} form a subnetwork of $N_{11}$, but this is not a one-port. Also, without loss of generality, we will only consider networks which are \emph{biconnected} (see \ref{nl:gt1}).

\begin{lemma}
\label{lem:pe}
Let $N$ be an RLC network with impedance $H(s)$, and let $N$ comprise the one-ports $\hat{N}_{1}, \ldots , \hat{N}_{m}$. Consider a sinusoidal trajectory of $N$ at an arbitrary but fixed frequency $\omega \in \mathbb{R}$. Denote the impedance of the one-port $\hat{N}_{k}$ by $Z_{k}(s)$, its phasor current by $\tilde{i}_{k}$, and its phasor voltage by $\tilde{v}_{k}$ ($k = 1, \ldots , m$). Then the following hold:
\begin{enumerate}[label=\arabic*., ref=\arabic*]
\item If $Z_{k}(s)$ has a pole at $s = j\omega$, then $\tilde{i}_{k} = 0$, otherwise $\tilde{v}_{k} = Z_{k}(j\omega)\tilde{i}_{k}$. \label{nl:feb12}
\item Suppose that either (i) $H(s)$ has a pole at $s = j \omega$, or (ii) $\Re\left(H(j\omega)\right) = 0$. If $Z_{k}(s)$ does not have a pole at $s = j\omega$ and $\Re{(Z_{k}(j\omega))} \neq 0$, then $\tilde{i}_{k} = \tilde{v}_{k} = 0$. \label{nl:feb21}
\end{enumerate}
\end{lemma}

\begin{IEEEproof}
To show condition \ref{nl:feb12}, we let $i_{k}(t) = \Re{(\tilde{i}_{k}e^{j\omega t})}$ and $v_{k}(t) = \Re{(\tilde{v}_{k}e^{j\omega t})}$ for all $t \in \mathbb{R}$. Then $\text{col}\!\begin{pmatrix}i_{k} &  \hspace{-0.15cm} v_{k}\end{pmatrix}$ is a sinusoidal driving-point trajectory (at frequency $\omega$) for the one-port $\hat{N}_{k}$, so condition \ref{nl:feb12} follows from note \ref{nl:gt8}.

For condition \ref{nl:feb21}, we denote the phasor current and voltage for the source by $\tilde{i}$ and $\tilde{v}$, respectively. From \ref{nl:gt3}, we obtain
\begin{equation}
\tilde{v}^{*}\tilde{i} + \tilde{i}^{*}\tilde{v} = \sum_{k = 1}^{m} \tilde{v}_{k}^{*} \tilde{i}_{k} + \tilde{i}_{k}^{*} \tilde{v}_{k}. \label{eq:s0rpe}
\end{equation}
From note \ref{nl:gt8}, if $H(s)$ has a pole at $s = j\omega$ (resp., $H(j\omega) = 0$), then $\tilde{i} = 0$ (resp., $\tilde{v} = 0$), and so $\tilde{v}^{*}\tilde{i} = \tilde{i}^{*}\tilde{v} = 0$. Otherwise, $\tilde{v}^{*}\tilde{i} + \tilde{i}^{*}\tilde{v} = \Re\left(H(j \omega)\right) \abs{\tilde{i}}^{2}$. Hence, the left-hand side of (\ref{eq:s0rpe}) is zero if either $H(s)$ has a pole at $s = j\omega$, or $\Re\left(H(j \omega)\right) = 0$.  Furthermore, if $Z_{k}(s)$ has a pole at $s = j \omega$, then $\tilde{i}_{k} = 0$ by condition \ref{nl:feb12}, and so $\tilde{v}_{k}^{*} \tilde{i}_{k} = \tilde{i}_{k}^{*} \tilde{v}_{k} = 0$. Otherwise, $\tilde{v}_{k}^{*} \tilde{i}_{k} + \tilde{i}_{k}^{*} \tilde{v}_{k} = \Re\left(Z_{k}(j\omega)\right) \abs{\tilde{i}_{k}}^{2}$, which is non-negative since $Z_{k}(s)$ is PR. Since all terms in the summation in (\ref{eq:s0rpe}) are non-negative then they must all be zero in order that their sum is zero. Hence, if $Z_{k}(s)$ does not have a pole at $s = j\omega$ and $\Re\left(Z_{k}(j\omega)\right) \neq 0$, then $\tilde{i}_{k} = 0$, so $\tilde{v}_{k} = Z_{k}(j\omega)\tilde{i}_{k} = 0$ by condition \ref{nl:feb12}. 
\end{IEEEproof}

From the above lemma, if $N$ is an RLC network which realizes a minimum function, then for any sinusoidal trajectory of $N$ at the minimum frequency $\omega_{0}$ there can be no current through or voltage across the resistors in $N$. We call the resistors \emph{blocked}, in accordance with the following definition.
\begin{definition}
\label{def:cvmb}
Let $N$ be an RLC network and let $\hat{n}$ be a (not necessarily one-port) subnetwork of $N$. For an arbitrary given trajectory of $N$, we call $\hat{n}$ \emph{blocked} if both the current through and the voltage across all of the elements in $\hat{n}$ are identically zero, and unblocked otherwise. We call $\hat{n}$ a \emph{maximal-blocked subnetwork} of $N$ if it is blocked and it is not contained within any larger blocked subnetwork of $N$.
\end{definition}

We now describe the structure of any RLC network whose impedance has a minimum frequency $\omega_{0}$ in terms of its maximal-blocked subnetworks and unblocked elements (with respect to a sinusoidal trajectory at frequency $\omega_{0}$). 
In the following lemma, we let $\omega_{0} > 0$ be fixed but arbitrary (and consistent with the $\omega_{0}$ in Appendix \ref{sec:nc}). 
\begin{lemma}
\label{lem:l5reis}
Let $N$ be an RLC network with impedance $H(s)$ which is not lossless, does not have a pole at $s = j\omega_{0}$, and satisfies $\Re \left(H(j\omega_{0})\right) = 0$ and $\Im \left(H(j\omega_{0})\right) \neq 0$. There exists a sinusoidal trajectory of $N$ at frequency $\omega_{0}$ for which the corresponding driving-point trajectory is non-zero. Moreover, for any such trajectory, the following hold: 
\begin{enumerate}[label=\arabic*., ref=\arabic*]
\item Neither the driving-point current nor the driving-point voltage are identically zero. \label{nl:gtnmf5rn1a}
\item All resistors in $N$ are blocked. \label{nl:gtnmf5rn1}
\item If the source is incident with a vertex in a maximal-blocked subnetwork of $N$, then an unblocked element is also incident with this vertex. \label{nl:gtnmf5rn2a}
\item If an unblocked element is incident with a vertex in a maximal-blocked subnetwork of $N$, then either (i) the source, or (ii) a second unblocked element, is also incident with this vertex. \label{nl:gtnmf5rn2ai}
\item Neither (i) the source, nor (ii) an unblocked element, can be incident with two vertices of the same maximal-blocked subnetwork of $N$. \label{nl:gtnmf5rn2b}
\end{enumerate}
If, in addition, $N$ contains four or fewer energy storage elements, then the following hold: 
\begin{enumerate}[label=\arabic*., ref=\arabic*]
\setcounter{enumi}{5}
\item There are either three or four unblocked elements in $N$, and each of these is an energy storage element. \label{nl:gtnmf5rnb3}
\item There are either one or two maximal-blocked subnetworks of $N$, and each of these subnetworks is a one-port.\label{nl:gtnmf5rnb4} 
\item Let $\hat{n}_{a}$ be one of the maximal-blocked one-ports in $N$. Then by either shorting or opening $\hat{n}_{a}$ in $N$ we obtain an RLC network $N_{a}$ whose impedance $H_{a}(s)$ satisfies $H_{a}(j\omega_{0}) = H(j\omega_{0})$. Also, when applicable, let $\hat{n}_{b}$ be the second maximal-blocked one-port in $N$, and suppose $\hat{n}_{b}$ is also a one-port in $N_{a}$. Then by either shorting or opening $\hat{n}_{b}$ in $N_{a}$ we obtain an RLC network $N_{b}$ whose impedance $H_{b}(s)$ satisfies $H_{b}(j\omega_{0}) = H(j\omega_{0})$.\label{nl:gtnmf5rnb4b}
\end{enumerate}
\end{lemma}

\begin{IEEEproof}
From \ref{nl:gt8}, there exists a sinusoidal trajectory of $N$ at frequency $\omega_{0}$ with a non-zero driving-point trajectory. We consider any such trajectory; we denote the impedance of $\hat{N}_{k}$ by $Z_{k}(s)$, its phasor current by $\tilde{i}_{k}$, and its phasor voltage by $\tilde{v}_{k}$ ($k = 1, \ldots , m$); and we denote the phasor current and voltage of the source by $\tilde{i}$ and $\tilde{v}$, respectively.

{\bf Proof of \ref{nl:gtnmf5rn1a}} \hspace{0.3cm} Since $H(s)$ does not have a pole at $s = j\omega_{0}$, $H(j\omega_{0}) \neq 0$, and $\text{col}\!\begin{pmatrix}\tilde{i}& \hspace{-0.15cm} \tilde{v}\end{pmatrix} \neq 0$, then it follows from note \ref{nl:gt8} that $\tilde{i} \neq 0$ and $\tilde{v} \neq 0$. 

{\bf Proof of \ref{nl:gtnmf5rn1}} \hspace{0.3cm} If the element $\hat{N}_{k}$ is a resistor, then $Z_{k}(s)$ does not have a pole at $s = j\omega_{0}$ and $\Re\left(Z_{k}(j\omega_{0})\right) \neq 0$, hence $\tilde{i}_{k} = \tilde{v}_{k} = 0$ by Lemma \ref{lem:pe}.

{\bf Proof of \ref{nl:gtnmf5rn2a}} \hspace{0.3cm} Let $\hat{n}$ be a maximal-blocked subnetwork in $N$, and let $x_{a}$ be a vertex in $\hat{n}$. If the source is incident with $x_{a}$, but no unblocked elements are incident with $x_{a}$, then $\tilde{i} = 0$ by Kirchhoff's current law. This contradicts condition \ref{nl:gtnmf5rn1a}. 

{\bf Proof of \ref{nl:gtnmf5rn2ai}} \hspace{0.3cm} Let $\hat{n}$ and $x_{a}$ be as in the proof of \ref{nl:gtnmf5rn2a}. If only one unblocked element $\hat{N}_{k}$ is incident with $x_{a}$, and the source is not incident with $x_{a}$, then $\tilde{i}_{k} = 0$. As $\hat{N}_{k}$ is a resistor, inductor, or capacitor, then $Z_{k}(s)$ does not have a pole at $s = j\omega_{0}$, so $\tilde{v}_{k} = Z_{k}(j\omega_{0})\tilde{i}_{k} = 0$ by Lemma \ref{lem:pe}. Thus, $\hat{n}$ is not a maximal-blocked subnetwork: a contradiction. 

{\bf Proof of \ref{nl:gtnmf5rn2b}} \hspace{0.3cm} Let $\hat{n}$ be a maximal-blocked subnetwork in $N$, and suppose either (i) the source, or (ii) a single unblocked element $\hat{N}_{k}$, is incident with two vertices $x_{a}$ and $x_{b}$ in $\hat{n}$. Since $\hat{n}$ is connected, then there is a path between $x_{a}$ and $x_{b}$ in $\hat{n}$. In case (i), by considering Kirchhoff's voltage law for the circuit comprised of the union of this path and the source, it is evident that $\tilde{v} = 0$, which contradicts condition \ref{nl:gtnmf5rn1a}. Similarly, in case (ii), we find that $\tilde{v}_{k} = 0$. Similar to the proof of \ref{nl:gtnmf5rn2ai}, since $\hat{N}_{k}$ is either a resistor, capacitor, or inductor, and $\tilde{v}_{k} = 0$, we conclude that $\tilde{i}_{k} = 0$, so $\hat{n}$ is not a maximal-blocked subnetwork: a contradiction. 

{\bf Proof of \ref{nl:gtnmf5rnb3}} \hspace{0.3cm} Since $H(s)$ is not lossless, then $N$ must contain at least one resistor \cite[Section III]{HugSmSP}. It then follows from condition \ref{nl:gtnmf5rn1} that there is at least one maximal-blocked subnetwork in $N$. Now, let $\hat{n}$ be a maximal-blocked subnetwork, and suppose $\hat{n}$ contains exactly $p$ vertices at which either (i) the source, or (ii) an unblocked element, is incident. It follows from conditions \ref{nl:gtnmf5rn1a}--\ref{nl:gtnmf5rn2b} that there must be at least $2p-1$ unblocked elements in $N$, and all of these are energy storage elements. Since $N$ is biconnected (see note \ref{nl:gt1}), then $p \geq 2$. Moreover, if there are four or fewer energy storage elements in $N$, then we require $p = 2$, implying that there must be either three or four unblocked elements in $N$.

{\bf Proof of \ref{nl:gtnmf5rnb4}} \hspace{0.3cm} Suppose there are exactly $q$ maximal-blocked subnetworks $\hat{n}_{1}, \ldots, \hat{n}_{q}$ in $N$. From the proof of \ref{nl:gtnmf5rnb3}, there are exactly two vertices of $\hat{n}_{k}$ at which unblocked elements are incident ($k = 1, \ldots , q$). This implies that each maximal-blocked subnetwork of $N$ is a one-port. Since, in addition, any vertex in $N$  is in at most one maximal-blocked subnetwork, then there are exactly $2q$ vertices in the maximal-blocked subnetworks of $N$ at which unblocked elements are incident. Furthermore, from conditions \ref{nl:gtnmf5rn2a} and \ref{nl:gtnmf5rn2ai}, then each of these vertices is incident with either 2 unblocked elements or 1 unblocked element and the source. As each element (and the source) is incident with exactly two vertices, then there must be at least $2q-1$ unblocked elements in $N$. Thus, if there are four or fewer energy storage elements in $N$, then $q \leq 2$.

{\bf Proof of \ref{nl:gtnmf5rnb4b}} \hspace{0.3cm} From \ref{nl:gt7+}, the network $N_{a}$ (and $N_{b}$ when applicable) has a sinusoidal trajectory at frequency $\omega_{0}$ whose driving-point trajectory is the same as the driving-point trajectory of $N$ considered in this proof. Since this driving-point trajectory is non-zero, then condition \ref{nl:gtnmf5rnb4b} follows from \ref{nl:gt8}. 
\end{IEEEproof}

In \cite{HugSmSP}, the realization of minimum functions using series-parallel networks was considered. We now generalize \cite[Theorem 5]{HugSmSP}. 
\begin{lemma}
\label{lem:mfnscpc}
Let $N$ be an RLC network which contains four or fewer energy storage elements, and let the impedance of $N$ be a minimum function. Then $N$ cannot be a series connection, nor a parallel connection, of two RLC networks.
\end{lemma}

\begin{IEEEproof}
Suppose that $N$ is a series connection of the RLC networks $\hat{N}_{1}$ and $\hat{N}_{2}$. Then $H(s) = Z_{1}(s) + Z_{2}(s)$ where $H(s), Z_{1}(s)$, and $Z_{2}(s)$ are the impedances of $N, \hat{N}_{1}$, and $\hat{N}_{2}$, respectively (see note \ref{nl:gt6}). Now, let $\omega_{0}$ be a minimum frequency. Then neither $Z_{1}(s)$ nor $Z_{2}(s)$ have any poles on $j \mathbb{R} \cup \infty$, and $\Re \left(Z_{1}(j\omega_{0})\right) = \Re \left(Z_{2}(j\omega_{0})\right) = 0$ \cite[Lemma 1]{HugSmSP}. In particular, neither $Z_{1}(s)$ nor $Z_{2}(s)$ is lossless, so both $\hat{N}_{1}$ and $\hat{N}_{2}$ contain at least two energy storage elements \cite[Lemma 2]{HugSmSP}. Since $\Im \left(H(j\omega_{0})\right) \neq 0$ then either $\Im \left(Z_{1}(j\omega_{0})\right) \neq 0$ or $\Im \left(Z_{2}(j\omega_{0})\right) \neq 0$, so either $\hat{N}_{1}$ or $\hat{N}_{2}$ contain at least three energy storage elements by Lemma \ref{lem:l5reis}. Hence, $N$ contains at least five energy storage elements: a contradiction. 

The case where $N$ is a parallel connection of two RLC networks is similar, and completes the proof.
\end{IEEEproof}

We now describe those RLC networks containing four or fewer energy storage elements which realize a minimum function. 

\begin{lemma}
\label{lem:mfffre}
Let $N$ be an RLC network containing four or fewer energy storage elements, and let the impedance of $N$ be a minimum function (with minimum frequency $\omega_{0}$). Then  $N$ comprises the one-ports $\hat{N}_{1}$ to $\hat{N}_{5}$ connected as in Fig.\ \ref{fig:ns1}, and $\hat{N}_{1}$ to $\hat{N}_{5}$ satisfy at least one of the following conditions.
\begin{enumerate}[label=\arabic*., ref=\arabic*]
\item $\hat{N}_{1}$ contains only resistors, $\hat{N}_{2}$ and $\hat{N}_{3}$ are both capacitors (resp., inductors), $\hat{N}_{4}$ and $\hat{N}_{5}$ are both inductors (resp., capacitors), and $Z_{2}(j\omega_{0})(Z_{3}(j\omega_{0})+Z_{4}(j\omega_{0})) + Z_{4}(j\omega_{0})(Z_{3}(j\omega_{0})+Z_{5}(j\omega_{0}))=0$. \label{nl:nc3}
\item $\hat{N}_{1}$ and $\hat{N}_{2}$ are both capacitors, $\hat{N}_{3}$ contains only resistors, $\hat{N}_{4}$ and $\hat{N}_{5}$ are both inductors, and $Z_{1}(j\omega_{0}) Z_{2}(j\omega_{0}) = Z_{4}(j\omega_{0}) Z_{5}(j\omega_{0})$ where $Z_{1}(j\omega_{0}) \neq -Z_{4}(j\omega_{0})$ and $Z_{1}(j\omega_{0}) \neq -Z_{5}(j\omega_{0})$. \label{nl:nc4}
\item $\hat{N}_{1}$ contains resistors together with at most one energy storage element, $\hat{N}_{2}$ contains only resistors, $\hat{N}_{3}$ contains only capacitors (resp., inductors), $\hat{N}_{4}$ and $\hat{N}_{5}$ contain only inductors (resp., capacitors), and $Z_{3}(j\omega_{0}) = -Z_{4}(j\omega_{0}) = -Z_{5}(j\omega_{0})$. \label{nl:nc1}
\item $\hat{N}_{1}$ and $\hat{N}_{2}$ each contain only resistors, $\hat{N}_{3}$ is a capacitor (resp., inductor), $\hat{N}_{4}$ comprises a series or parallel connection of an inductor and capacitor, $\hat{N}_{5}$ is an inductor (resp., capacitor), and $Z_{3}(j\omega_{0}) {=} -Z_{4}(j\omega_{0}) {=} -Z_{5}(j\omega_{0})$.\label{nl:nc2}
\end{enumerate}
\end{lemma}

\begin{figure}[!t]
\scriptsize
\begin{center}
\includegraphics[page=22,width=0.5\hsize]{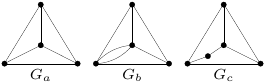}
\end{center}
\caption{Three biconnected graphs with seven or fewer edges.}
\label{fig:cg4v}
\end{figure}

\begin{figure}[t]
\scriptsize
\begin{center}
\includegraphics[page=23,width=0.62\hsize]{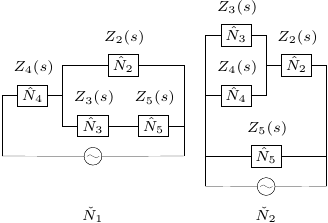}
\end{center}
\caption{Networks $\check{N}_{1}$ and $\check{N}_{2}$ described in the proof of Lemma \ref{lem:mfffre}. $\check{N}_{1}$ (resp., $\check{N}_{2}$) is obtained by opening (resp., shorting) $\hat{N}_{1}$ in the network $N$ in Fig.\ \ref{fig:ns1}. Here, $\check{H}_{1} = ((Z_{3}+Z_{5})(Z_{2}+Z_{4})+Z_{2}Z_{4})/(Z_{2}+Z_{3}+Z_{5})$, and $\check{H}_{2} = Z_{5}(Z_{2}Z_{3}+Z_{2}Z_{4}+Z_{3}Z_{4})/((Z_{3}+Z_{4})(Z_{2}+Z_{5})+Z_{3}Z_{4})$, where $\check{H}_{1}$ and $\check{H}_{2}$ denote the impedances of $\check{N}_{1}$ and $\check{N}_{2}$, respectively.}
\label{fig:n1id}
\end{figure}

\begin{figure}[t]
\scriptsize
\begin{center}
\includegraphics[page=24,width=0.5\hsize]{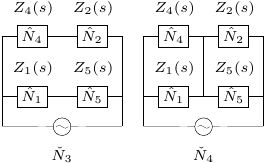}
\end{center}
\caption{Networks $\check{N}_{3}$ and $\check{N}_{4}$ described in the proof of Lemma \ref{lem:mfffre}. $\check{N}_{3}$ (resp., $\check{N}_{4}$) is obtained by opening (resp., shorting) $\hat{N}_{3}$ in the network $N$ in Fig.\ \ref{fig:ns1}. Here, $\check{H}_{3} = (Z_{1}+Z_{5})(Z_{2}+Z_{4})/(Z_{1}+Z_{2}+Z_{4}+Z_{5})$, and $\check{H}_{4} = (Z_{1}Z_{4}(Z_{2}+Z_{5})+Z_{2}Z_{5}(Z_{1}+Z_{4}))/((Z_{1}+Z_{4})(Z_{2}+Z_{5}))$, where $\check{H}_{3}$ and $\check{H}_{4}$ denote the impedances of $\check{N}_{3}$ and $\check{N}_{4}$, respectively.}
\label{fig:n2id}
\end{figure}

\begin{figure}[!t]
\scriptsize
\begin{center}
\includegraphics[page=25,width=0.5\hsize]{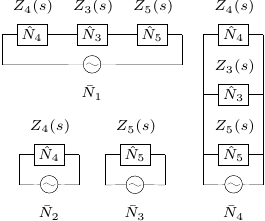}
\end{center}
\caption{Networks $\bar{N}_{1}$, $\bar{N}_{2}$, $\bar{N}_{3}$, and $\bar{N}_{4}$ described in the proof of Lemma \ref{lem:mfffre}. $\bar{N}_{1}$ (resp., $\bar{N}_{2}$) is obtained by opening (resp., shorting) $\hat{N}_{2}$ in the network $\check{N}_{1}$ in Fig.\ \ref{fig:n1id}; and $\bar{N}_{3}$, (resp., $\bar{N}_{4}$) is obtained by opening (resp., shorting) $\hat{N}_{2}$ in the network $\check{N}_{2}$ in Fig.\ \ref{fig:n1id}. Here, $\bar{H}_{1} = Z_{3}+Z_{4}+Z_{5}$, $\bar{H}_{2} = Z_{4}$, $\bar{H}_{3} = Z_{5}$, and $\bar{H}_{4} = Z_{3}Z_{4}Z_{5}/(Z_{3}Z_{4}+Z_{3}Z_{5}+Z_{4}Z_{5})$, where $\bar{H}_{1}, \bar{H}_{2}, \bar{H}_{3}$, and $\bar{H}_{4}$ denote the impedances of $\bar{N}_{1}, \bar{N}_{2}, \bar{N}_{3}$, and $\bar{N}_{4}$, respectively.}
\label{fig:n3id}
\end{figure}

\begin{IEEEproof}
Let $H(s)$ denote the impedance of $N$. Since $H(s)$ is a minimum function, then it satisfies the conditions in Lemma \ref{lem:l5reis}. We will first show that $N$ comprises the one-ports $\hat{N}_{1}, \ldots , \hat{N}_{5}$ connected as in Fig.\ \ref{fig:ns1}, and one of the following two conditions holds: 
\begin{enumerate}[label=(\alph*)]
\item Exactly one of the one-ports $\hat{N}_{1}, \ldots, \hat{N}_{5}$ corresponds to a maximal-blocked subnetwork, and the remaining one-ports are comprised of energy storage elements.\label{nl:lgmfsca}
\item Exactly two of the one-ports $\hat{N}_{1}, \ldots, \hat{N}_{5}$ correspond to maximal-blocked subnetworks, these two one-ports are not incident with the same vertex, and the remaining one-ports are comprised of energy storage elements.\label{nl:lgmfscb}
\end{enumerate}
To show this, we note initially that $N$ comprises either one or two maximal-blocked subnetworks which are one-ports, and either three or four unblocked elements which are all energy storage elements, by Lemma \ref{lem:l5reis}. Now, let $G$ denote the associated graph of $N$  (we recall that $G$ includes an edge corresponding to the source). Then $G$ contains at most seven edges, is biconnected, and, by Lemma \ref{lem:mfnscpc}, it must not correspond to a series or parallel connection of two RLC networks. By \cite[p.\ 326]{Telge}, it must be either the complete graph on four vertices (graph $G_{a}$ in Fig.\ \ref{fig:cg4v}), or the graph obtained by replacing any single edge in this graph by either two edges in series or two in parallel (graphs $G_{b}$ and $G_{c}$ in Fig.\ \ref{fig:cg4v}). By Lemma \ref{lem:mfnscpc}, if $G$ takes the form of $G_{b}$ (resp., $G_{c}$), then neither of the edges connected in series (resp., parallel) can correspond to the source. Thus, irrespective of whether $G$ takes the form of $G_{a}, G_{b}$ or $G_{c}$, we find that $N$ comprises the one-ports $\hat{N}_{1}, \ldots , \hat{N}_{5}$ connected as in Fig.\ \ref{fig:ns1}. Also, from Lemma \ref{lem:l5reis}, then either \ref{nl:lgmfsca} or \ref{nl:lgmfscb} must hold.

To complete the proof, we will show that condition \ref{nl:nc3} or \ref{nl:nc4} (resp., \ref{nl:nc1} or \ref{nl:nc2}) must hold in case \ref{nl:lgmfsca} (resp., \ref{nl:lgmfscb}). To show this, we note initially that $H(s)$ has no poles or zeros on $j\mathbb{R} \cup \infty$, so $N$ must not contain a driving-point L-cut-set, C-cut-set, L-path or C-path (see \ref{nl:gt9}). 

{\bf Case \ref{nl:lgmfsca}} \hspace{0.3cm} Without loss of generality, we can let the blocked subnetwork be either $\hat{N}_{1}$ or $\hat{N}_{3}$ (this follows from \ref{nl:gt2}). Furthermore, the remaining subnetworks must each comprise a single energy storage element. 

Suppose initially that $\hat{N}_{1}$ corresponds to the blocked one-port. Then, with $\check{H}_{1}(s)$ and $\check{H}_{2}(s)$ as in Fig.\ \ref{fig:n1id}, we require $\check{H}_{1}(j\omega_{0}) = \check{H}_{2}(j\omega_{0})$ by condition \ref{nl:gtnmf5rnb4b} of Lemma \ref{lem:l5reis}. Since, in addition, $N$ must not contain a driving-point L-cut-set, C-cut-set, L-path or C-path, we find that condition \ref{nl:nc3} must hold. 

Next, suppose that $\hat{N}_{3}$ corresponds to the blocked one-port. Then, with $\check{H}_{3}(s)$ and $\check{H}_{4}(s)$ as in Fig.\ \ref{fig:n2id}, we require $\check{H}_{3}(j\omega_{0}) = \check{H}_{4}(j\omega_{0})$. As $N$ must not contain a driving-point L-cut-set, C-cut-set, L-path or C-path, we find that condition \ref{nl:nc4} must hold.

{\bf Case \ref{nl:lgmfscb}} \hspace{0.3cm}  Without loss of generality, we can let the blocked subnetworks be $\hat{N}_{1}$ and $\hat{N}_{2}$  (this again follows from \ref{nl:gt2}). Also, since there must be at least three energy storage elements which are not in these one-ports, and $N$ contains four or fewer energy storage elements, then at least one of these two one-ports must contain only resistors. Hence, without loss of generality, we can let $\hat{N}_{2}$ contain only resistors. Now, with $\bar{H}_{1}(s), \bar{H}_{2}(s), \bar{H}_{3}(s)$, and $\bar{H}_{4}(s)$ as in Fig.\ \ref{fig:n3id}, we require $\bar{H}_{1}(j\omega_{0}) = \bar{H}_{2}(j\omega_{0}) = \bar{H}_{3}(j\omega_{0}) = \bar{H}_{4}(j\omega_{0})$ by Lemma \ref{lem:l5reis}. This implies 
\begin{equation}
\label{eq:iegmfn1}
Z_{3}(j\omega_{0}) = -Z_{4}(j\omega_{0}) = -Z_{5}(j\omega_{0}).
\end{equation}
There are now three cases to consider: (i) $\hat{N}_{3}$, $\hat{N}_{4}$ and $\hat{N}_{5}$ each comprise one energy storage element, and $\hat{N}_{1}$ contains resistors and at most one energy storage element; (ii) $\hat{N}_{3}$ comprises two energy storage elements, $\hat{N}_{4}$ and $\hat{N}_{5}$ each comprise one energy storage element, and $\hat{N}_{1}$ contains only resistors; and (iii) $\hat{N}_{4}$ comprises two energy storage elements, $\hat{N}_{3}$ and $\hat{N}_{5}$ each comprise one energy storage element, and $\hat{N}_{1}$ contains only resistors.

In case (i), equation (\ref{eq:iegmfn1}) implies that $\hat{N}_{4}$ and $\hat{N}_{5}$ must comprise energy storage elements of the same type, and of opposite type to $\hat{N}_{3}$, and so condition \ref{nl:nc1} holds.

In case (ii), equation (\ref{eq:iegmfn1}) implies that $\hat{N}_{4}$ and $\hat{N}_{5}$ comprise energy storage elements of the same type. If $\hat{N}_{3}$ comprises an inductor and a capacitor, then $\hat{N}_{3}$ contains a driving-point L-cut-set, C-cut-set, L-path or C-path, so $\hat{N}_{3}$ must contain only one type of energy storage element. From equation (\ref{eq:iegmfn1}), this energy storage element is of opposite type to those in $\hat{N}_{4}$ and $\hat{N}_{5}$. Thus, condition \ref{nl:nc1} also holds in this case.

In case (iii), equation (\ref{eq:iegmfn1}) implies that $\hat{N}_{3}$ and $\hat{N}_{5}$ comprise energy storage elements of opposite types, and the energy storage elements in $\hat{N}_{4}$ cannot all be of the same type as $\hat{N}_{3}$. It is then clear that either condition \ref{nl:nc1} or \ref{nl:nc2} holds. 
\end{IEEEproof}

Note that each of the one-ports in Lemma \ref{lem:mfffre} contain at most two types of elements. 
Using established results concerning such networks, we now prove Theorems \ref{thm:mfr34re} and \ref{thm:mfr4re}.
 
\begin{IEEEproof}[Proof of Theorem \ref{thm:mfr34re}]
The impedance of any RLC network containing only one type of element is equivalent to the impedance of a single element  of that type. Theorem \ref{thm:mfr34re} then follows from \ref{nl:gt3b} and Lemma \ref{lem:mfffre}, noting that condition \ref{nl:nc1} in that lemma must hold if $N$ contains exactly three energy storage elements.
\end{IEEEproof}

\begin{IEEEproof}[Proof of Theorem \ref{thm:mfr4re}]
It is well known that the impedance of any RLC network which contains only two types of element is equivalent to the impedance of one of the \emph{Cauer canonical networks} (see, e.g., \cite{HugCF}). Theorem \ref{thm:mfr4re} then follows from \ref{nl:gt3b} and Lemma \ref{lem:mfffre}.
\end{IEEEproof}

\section{RLC realizations of biquadratic minimum functions}
\label{sec:rbmfrlc}
In this section, we prove Theorems \ref{thm:bmfr3re} and \ref{thm:bmfr4re}. This involves determining which of the networks in Lemma \ref{lem:mfffre} realize biquadratic minimum functions. The networks in Figs.\ \ref{fig:mfrn3} and \ref{fig:mfrn4} are eliminated by the following lemma.

\begin{lemma}
\label{lem:bmf2r}
Let $N$ be an RLC network whose impedance is a biquadratic minimum function. Then the resistors in $N$ cannot all be contained in a single one-port subnetwork of $N$ comprised of resistors alone.
\end{lemma}

\begin{IEEEproof}
Let $H(s)$ denote the impedance of $N$, Then $H(s)$ has neither a pole nor a zero at $s = 0$ or $s = \infty$. Also, $H(s)$ takes the form indicated in Lemma \ref{lem:bmfp}, whence $W > 0$ and $W \neq 1$, and so $H(0) \neq H(\infty)$. 

Now, suppose all the resistors in $N$ are contained in a single one-port $\hat{N}_{k}$ comprised of resistors alone. Then the impedance of $\hat{N}_{k}$ is equal to some positive constant $R$, and the impedance of $N$ is the same as the network $N_{a}$ obtained by replacing the one-port $\hat{N}_{k}$ in $N$ with a single resistor $\hat{N}$ of resistance $R$ (see note \ref{nl:gt3b}). By applying the results in \ref{nl:gt9} inductively to open all the capacitors and short all the inductors in $N_{a}$, we obtain a network $N_{b}$ which is either (i) the resistor $\hat{N}$, (ii) an open circuit, or (iii) a short circuit. Again from \ref{nl:gt9}, since $H(s)$ does not have a pole or zero at $s = 0$, then $N_{b}$ must be the resistor $\hat{N}$, and $H(0) = R$. Similarly, by considering the network obtained by shorting all the inductors and opening all the capacitors in $N_{a}$, we conclude that $H(\infty) = R$. This implies $H(0) = H(\infty)$: a contradiction.
\end{IEEEproof}

Using an algebraic argument, we now prove Theorem \ref{thm:bmfr3re}. In this proof, we use the notation $R_{k}(p(s), q(s))$ to denote the \emph{Sylvester determinants} of two polynomials $p(s) = p_{m}s^{m} + p_{m-1}s^{m-1} + \cdots$ and $q(s) = q_{n}s^{n} + q_{n-1}s^{n-1} + \cdots$, where $p_{m} \neq 0$ and $q_{n} \neq 0$: 
\begin{equation*}
R_{k}(p(s),q(s)) \coloneqq \begin{matrix} 
\left\lvert\begin{smallmatrix}
p_{m} & p_{m-1} & \cdots & & \\
0 & p_{m} & p_{m-1} & \cdots & \\
& & \ddots & & \\
q_{n} & q_{n-1} & \cdots & & \\
0 & q_{n} & q_{n-1} & \cdots & \\
& & \ddots & & \\
\end{smallmatrix}\right\rvert 
\mspace{-10mu}
&
\begin{matrix}    
\left. \vphantom{\begin{smallmatrix} \cdots \\ p_{m-1} \\ \vdots \end{smallmatrix}}\right\}
\text{$n-k$ rows} \\
\left. \vphantom{\begin{smallmatrix} \cdots \\ q_{n-1} \\ \vdots \end{smallmatrix}}\right\}
\text{$m-k$ rows} \\
\end{matrix}
\end{matrix},
\end{equation*}
for $k = 0, 1, \ldots, (\min{\lbrace m,n\rbrace} - 1)$. From \cite[Theorem 15]{HugSmSP}, $p(s)$ and $q(s)$ have at least $r$ roots in common (counting according to multiplicity) if and only if $R_{0}(p(s), q(s)) = \cdots = R_{r-1}(p(s), q(s)) = 0$.

\begin{IEEEproof}[Proof of Theorem \ref{thm:bmfr3re}]
\label{pr:mtp}
That $N$ contains at least three energy storage elements and at least two resistors follows from Theorem \ref{thm:mfr34re} and Lemma \ref{lem:bmf2r}. Since $H(s)$ is a biquadratic minimum function, then $H(s)$ takes the form indicated in Lemma \ref{lem:bmfp}. Also, by direct calculation, the impedances of $N_{1}$ and $N_{2}$ satisfy conditions (a) and (b), respectively. Now, let $N$ contain exactly three energy storage elements. It then follows from Theorem \ref{thm:mfr34re} and Lemma \ref{lem:bmf2r} that $H(s)$ is the impedance of a network from $\mathcal{Q}_{7}$ (see Fig.\ \ref{fig:mfrn1}). To complete the proof, we will show that the functions described in (a) and (b) are the only biquadratic minimum functions which can be realized as the impedance of a network from $\mathcal{Q}_{7}$.

We first consider those circumstances in which the impedance $H_{7}(s)$ of network $N_{7}$ in Fig.\ \ref{fig:mfrn1} takes the form indicated in Lemma \ref{lem:bmfp}. We thus require $H_{7}(j\omega_{0}) = KFj$, $H_{7}(0) = KW^{2}$, and $H_{7}(\infty) = K$. For $H_{7}(j\omega_{0}) = KFj$, we require $C = K F$, and hence $F>0$ which implies $0 < W < 1$. Let $g_{1} \coloneqq K/A$ and $g_{2} \coloneqq K/B$. Then, we require $g_{1}, g_{2} > 0$ and 
\begin{equation}
\hspace*{-0.3cm} \frac{1}{g_{1} + g_{2}} = \frac{H_{7}(0)}{K} = W^{2}, \hspace{0.1cm} \text{and } \frac{g_{1} + g_{2}}{g_{1}g_{2}} = \frac{H_{7}(\infty)}{K} = 1. \label{eq:qwiiz}
\end{equation}
Now, let $p(s)$ and $q(s)$ be the polynomials of degree three in $s$ such that $p(0) = \omega_{0}^{3}$ and $p(s)/q(s) = H_{7}(s)/(K F)$. It is easily verified that the terms in $p(s)$ and $q(s)$ of degree three in $s$ cannot be zero, so for $H_{7}(s)$ to be biquadratic we require $R_{0}(p(s), q(s)) = F^{2}\omega_{0}^{9}(g_{1} - g_{2})^{2}(1+F^{2}g_{1}g_{2})^{4} = 0$. Together with (\ref{eq:qwiiz}), and the conditions $F, \omega_{0}, g_{1}, g_{2} > 0$, this implies $g_{1} = g_{2} = 2$ and $W = 1/2$. Thus, if the impedance of a network from $\mathcal{N}_{7}$ is a biquadratic minimum function, then condition (a) holds.

By a duality argument (see Appendix \ref{sec:nc}) it is easily shown that if the impedance of a network from $\mathcal{N}_{7}^{d}$ is a biquadratic minimum function, then condition (b) holds. This completes the proof.
\end{IEEEproof}

\begin{IEEEproof}[Proof of Theorem \ref{thm:bmfr4re}]
\label{pr:mtp2}
Since $H(s)$ is a biquadratic minimum function, then it takes the form indicated in Lemma \ref{lem:bmfp}. Direct calculation verifies that  the impedances of networks $N_{3}, N_{4}, N_{5}$, and $N_{6}$ satisfy conditions \ref{nl:4rebmfc1}, \ref{nl:4rebmfc2}, \ref{nl:4rebmfc3}, and \ref{nl:4rebmfc4}, respectively. Now, let $N$ contain at most four energy storage elements. It then follows from Theorem \ref{thm:mfr34re} and Lemma \ref{lem:bmf2r} that $H(s)$ is the impedance of a network from one of the network classes in 
 Figs.\ \ref{fig:mfrn1}, \ref{fig:mfrn2}, and \ref{fig:mfrn5}. We will show that the functions described in conditions (a)--\ref{nl:4rebmfc4} of Theorems \ref{thm:bmfr3re} and \ref{thm:bmfr4re} are the only biquadratic minimum functions realized as the impedance of a network from one of these classes. The case of $N$ belonging to $\mathcal{Q}_{7}$ was considered in the proof of Theorem \ref{thm:bmfr3re}, so there are three remaining cases to consider:
\begin{enumerate}[label=(\roman*)]
\item $N$ belongs to $\mathcal{Q}_{8}$ (see Fig.\ \ref{fig:mfrn2}).\label{nl:bmfc1}
\item $N$ belongs to one of the classes $\mathcal{N}_{11}$, $\mathcal{N}_{11}^{i}$, $\mathcal{N}_{11a}$, $\mathcal{N}_{11a}^{i}$, $\mathcal{N}_{11b}$ or $\mathcal{N}_{11b}^{i}$ (see Fig.\ \ref{fig:mfrn5}).\label{nl:bmfc2}
\item $N$ belongs to one of the classes $\mathcal{N}_{12}$, $\mathcal{N}_{12}^{i}$, $\mathcal{N}_{12a}$, $\mathcal{N}_{12a}^{i}$, $\mathcal{N}_{12b}$ or $\mathcal{N}_{12b}^{i}$ (see Fig.\ \ref{fig:mfrn5}).\label{nl:bmfc3}
\end{enumerate}

{\bf Case \ref{nl:bmfc1}} \hspace{0.3cm}  Let the impedance $H_{8}(s)$ of network $N_{8}$ in Fig.\ \ref{fig:mfrn2} take the form indicated in Lemma \ref{lem:bmfp}. For $H_{8}(j\omega_{0}) = KFj$ we require $D = F > 0$, which implies $0<W<1$. Let $g_{1} \coloneqq K /A$, $g_{2} \coloneqq K/B$, and $c_{2} \coloneqq K F/C$, and so $g_{1}, g_{2}, c_{2} > 0$. Then we require
\begin{equation}
\frac{1}{g_{1} + g_{2}} = \frac{H_{8}(0)}{K} = W^{2}, \hspace{0.1cm} \text{and } \frac{1}{g_{2}} = \frac{H_{8}(\infty)}{K} = 1. \label{eq:q2iiz}
\end{equation}
Next, let $p(s)$ and $q(s)$ be the polynomials of degree four in $s$ such that $p(0) = (1+c_{2})\omega_{0}^{4}$ and $p(s)/q(s) = H_{8}(s)/(K F)$. We find that the terms in $p(s)$ and $q(s)$ of degree four in $s$ cannot be zero, so for $H_{8}(s)$ to be biquadratic we require $R_{0}(p(s),q(s)) = R_{1}(p(s),q(s)) = 0$. Here, $R_{0}(p(s),q(s)) = c_{2}\omega_{0}^{16}(1+c_{2})(1+F^{2}g_{1}g_{2})^{4}f_{1}^{2}$ and $R_{1}(p(s),q(s)) = -c_{2}\omega_{0}^{9}(1+F^{2}g_{1}g_{2})^{2}f_{2}$ where $f_{1}$ and $f_{2}$ are both polynomials in $c_{2}, g_{1}, g_{2}$ and $F$. We thus require $f_{1} = f_{2} = 0$, so, in particular, $R_{0}(f_{1}(F), f_{2}(F)) = c_{2}^{6}g_{2}^{10}(1+c_{2})^{2}(c_{2}^{2}g_{1}+2c_{2}(g_{1}-g_{2})+g_{1}-3g_{2})^{2} = 0$. Taken together with (\ref{eq:q2iiz}) and the conditions $c_{2}, g_{1}, g_{2} > 0$ and $0<W<1$, this implies $g_{1} = (1-W^{2})/W^{2}$, $g_{2} = 1$, $c_{2} = (2W-1)/(1-W)$, and $W > 1/2$. Then $R_{0}(p(s), q(s)) = 0$ and $F > 0$ imply $F = W\sqrt{2W-1}/(1-W)$, so condition \ref{nl:4rebmfc1} holds. It is then easily shown from duality and frequency inversion arguments (see Appendix \ref{sec:nc}) that, in case \ref{nl:bmfc1}, one of the conditions \ref{nl:4rebmfc1}--\ref{nl:4rebmfc4} must hold.

{\bf Case \ref{nl:bmfc2}} \hspace{0.3cm}  We let the impedance $H_{11}(s)$ of $N_{11}$ in Fig.\ \ref{fig:mfrn5} take the form indicated  in Lemma \ref{lem:bmfp}. For $H_{11}(j\omega_{0}) = KFj$ we require $E = F$, which implies $F > 0$ and $0 < W < 1$. Let $r_{1} \coloneqq A/K$, $g_{3} \coloneqq K / B$, $g_{2} \coloneqq KC$, and $c_{1} \coloneqq K F /D$, so $g_{3}, c_{1} > 0$ and $r_{1}, g_{2} \geq 0$. Then, by considering $H_{11}(0)$ and $H_{11}(\infty)$, we obtain 
\begin{equation}
\frac{1+r_{1}g_{2}}{g_{2}(1+r_{1}g_{3})+g_{3}} = W^{2} , \hspace{0.1cm} \text{and } \frac{1+r_{1}g_{3}}{g_{3}} = 1 \label{eq:q5iiz}.
\end{equation}
Let $p(s)$ and $q(s)$ be the polynomials of degree four in $s$ such that  $p(0) = F(1+r_{1}g_{2})\omega_{0}^{4}$ and $p(s)/q(s) = H_{11}(s)/ (K F)$. We find that the terms in $p(s)$ and $q(s)$ of degree four in $s$ cannot be zero, so for $H_{11}(s)$ to be biquadratic we require $R_{0}(p(s),q(s)) = R_{1}(p(s),q(s)) = 0$. In this case, $R_{0}(p(s), q(s)) = F^{4}\omega_{0}^{16}c_{1}(c_{1}^{2}(r_{1}+F^{2}g_{3})^{2}+F^{2}(1+r_{1}g_{2}+F^{2}g_{2}g_{3})^{2})^{2}f_{1}^{2}$ and $R_{1}(p(s), q(s)) = -F^{2}c_{1}\omega_{0}^{9}f_{2}$ where $f_{1}$ and $f_{2}$ are both polynomials in $c_{1}, r_{1}, g_{2}, g_{3}$, and $F$. We thus require $R_{0}(f_{1}(c_{1}), f_{2}(c_{1})) = F^{10}g_{3}^{5}(r_{1}g_{3}-1)((g_{2}(1-r_{1}g_{3})(r_{1}+F^{2}g_{3})+F^{2}g_{3}^{2}+g_{3}r_{1}-1)^{2}+g_{3}^{2}F^{2})^{2}(g_{3}(3-r_{1}g_{3}+r_{1}g_{2}(2-r_{1}g_{3}))-g_{2}) = 0$, and $f_{1} = c_{1}(1-r_{1}g_{3}) + F^{2}g_{3}(g_{3} - g_{2}(1-r_{1}g_{3})) = 0$. Taken together with (\ref{eq:q5iiz}), and the conditions $r_{1}, g_{2} \geq 0$, $g_{3}, c_{1} > 0$, this implies $r_{1} = (g_{3}-1)/g_{3}$, $g_{2} = g_{3}(4-g_{3})/(g_{3}-2)^{2}$, $c_{1} = 2F^{2}g_{3}^{2}/ (2-g_{3})^{2}$, $W = 1/2$, and $1 \leq g_{3} \leq 4$, $g_{3} \neq 2$, so condition (a) of Theorem \ref{thm:bmfr3re} holds. Thus, we conclude by duality and frequency inversion arguments that, in case \ref{nl:bmfc2}, one of the conditions (a) or (b) in Theorem \ref{thm:bmfr3re} must hold. 

{\bf Case \ref{nl:bmfc3}} \hspace{0.3cm}  Let the impedance $H_{12}(s)$ of network $N_{12}$ in Fig.\ \ref{fig:mfrn5} take the form indicated  in Lemma \ref{lem:bmfp}. In this case, $H_{12}(j\omega_{0}) = KFj$ implies $E=F$, and so $F > 0$ and $0 < W < 1$. Now, let $r_{1} \coloneqq A/K$, $g_{2}\coloneqq  KC$, $g_{3} \coloneqq K /B$, and $x_{1} \coloneqq K F /D$, so $g_{3}, x_{1} > 0$ and $r_{1}, g_{2} \geq 0$. Then let $p(s)$ and $q(s)$ be the polynomials of degree $4$ in $s$ such that  $p(0) = r_{1}x_{1}\omega_{0}^{4}$ and $p(s)/q(s) = H_{12}(s)/(K F)$. We find that the terms in $p(s)$ and $q(s)$ of degree four in $s$ cannot be zero, and in this case we find $R_{0}(p(s),q(s)) = -F^{6}\omega_{0}^{16}x_{1}((F^{2}g_{3}+r_{1})^{2}x_{1}^{2}+F^{2}(1+r_{1}g_{2}+F^{2}g_{2}g_{3})^{2})^{2}f_{1}^{2}$ and $R_{1}(p(s),q(s)) = F^{6}\omega_{0}^{9}f_{2}$ where $f_{1}$ and $f_{2}$ are both polynomials in $x_{1}, r_{1}, g_{2}, g_{3}$, and $F$. We thus require $R_{0}(f_{1}(x_{1}), f_{2}(x_{1})) = -F^{2}g_{3}^{5}((g_{2}(1-r_{1}g_{3})(r_{1}+F^{2}g_{3}) - g_{3}r_{1})^{2}+g_{3}^{2}F^{2})^{2}(g_{3} - g_{2}(1-r_{1}g_{3}))(g_{2}(1-r_{1}g_{3})^{2}+g_{3}(1+r_{1}g_{3})) = 0$, together with $f_{1} = g_{3}(1-r_{1}g_{3})x_{1} + g_{3} - g_{2}(1-r_{1}g_{3}) = 0$. It may be verified that these equations have no solution for $r_{1}, g_{2} \geq 0$ and $g_{3}, x_{1}, F > 0$. It follows from duality and frequency inversion arguments that $H(s)$ cannot be biquadratic in case \ref{nl:bmfc3}: a contradiction.
\end{IEEEproof}

\begin{IEEEproof}[Proof of Theorem \ref{thm:bmfr5re}]
Condition \ref{nl:5recon1} follows from Theorems \ref{thm:bmfr3re} and \ref{thm:bmfr4re}. To see condition \ref{nl:5recon2}, first note from Theorem \ref{thm:rmf} and Lemma \ref{lem:bmfp} that $H(j\omega_{0}) = \omega_{0}Xj = KFj$, so $X = KF/\omega_{0}$. Thus, $F, \omega_{0}, K > 0$ imply $X > 0$, so $H(s)$ is realized as the impedance of the networks on the top left and bottom right of Figs.\ \ref{fig:ps} and \ref{fig:nabd} by Theorem \ref{thm:rmf}, where $\mu$, $H(\mu)$, $\alpha$ and $H_{r}(s)$ are as defined in that theorem. It it is then easily verified that when $H(s)$ takes the form indicated in Lemma \ref{lem:bmfp}, then $\mu$, $H(\mu)$, $\alpha$ and $H_{r}(s)$ are as in condition \ref{nl:5recon2}. Since $H(\mu)/H_{r}(s)$ and $H(\mu)H_{r}(s)$ are both positive constants, then both $\hat{N}_{1}$ and $\hat{N}_{2}$ can be replaced with a resistor. Condition \ref{nl:5recon3} may then be shown similarly.
\end{IEEEproof}

\section{Conclusions}
\label{sec:concl}
The networks discovered in the 1950s \cite{Reza_54, PS, Fialkow_Gerst}, which simplify the famous Bott-Duffin networks \cite{BD}, contain a surprisingly large number of energy storage elements, and have non-minimal state-space realizations with states as inductor currents and capacitor voltages. In this paper, we showed that these networks actually contain the least possible number of energy storage elements for realizing certain impedances (almost all biquadratic minimum functions). In particular, we proved six theorems on the realization of minimum functions with RLC networks. The main argument was summarised after Theorem \ref{thm:mfr4re}. It is based on the observation that, for an RLC network $N$ realizing a minimum function, and a sinusoidal trajectory of $N$ at the minimum frequency, there is no energy dissipated in $N$ over a single period, so only the energy storage elements can transmit current at this frequency.

\section*{Acknowledgments}
The author would like to thank R.A.V.\ Robison and M.C.\ Smith for many helpful discussions, and the anonymous reviewers for their valuable comments.

\appendices

\section{Network Classification}
\label{sec:nc}
Here, we provide some network classification terminology, which enables a concise presentation of our main results. 

As in \cite{HugSmSP}, we will define networks using diagrams within figures (e.g., $N_{1}$ in Fig.\ \ref{fig:mfrn1}), in which we indicate the driving-point terminals with dots, we write the impedance of an element above the element, and we list constraints on the element impedances in the figure's caption. Each diagram also defines a network class corresponding to the set of all networks of the type indicated whose element impedances satisfy the constraints listed in the corresponding caption.

The concepts of duality and frequency inversion in RLC network analysis were exploited in \cite{JiangSmith11, HugSmSP}. As in \cite{HugSmSP}, we let $\omega_{0} > 0$ be arbitrary but fixed, and we consider frequency inversion with  respect to $\omega_{0}$. If $H(s)$ is a minimum function with $\omega_{0}$ a minimum frequency, then so too are $1/H(s)$ and $H(\omega_{0}^{2}/s)$ \cite{HugSmSP}. In particular, consider the parametrisation of a minimum function described in Lemma \ref{lem:bmfp} as a function of $s, K, \omega_{0}, W,$ and $F$, i.e.,
\begin{equation*}
H(s, K, \omega_{0}, W, F) = K\frac{s^{2} + \frac{\omega_{0}\left(1-W\right)F}{W}s+\omega_{0}^{2}W}{s^{2} + \frac{\omega_{0}\left(1-W\right)}{F}s + \frac{\omega_{0}^{2}}{W}}.
\end{equation*}
Then we note the relationships:
\begin{align*}
H(\tfrac{\omega_{0}^{2}}{s},K,\omega_{0},W,F) = H(s, KW^{2}, \omega_{0}, \tfrac{1}{W}, -\tfrac{F}{W^{2}}),\\
\text{and } \tfrac{1}{H(s,K,\omega_{0},W,F)} = H(s, \tfrac{1}{K},\omega_{0},\tfrac{1}{W},-\tfrac{1}{F}).
\end{align*}

Now, let $N$ be an RLC network with impedance $H(s)$. Then $N$ has a \emph{frequency inverted} network $N^{i}$ whose impedance is $H(\omega_{0}^{2}/s)$ \cite{HugSmSP}. If, in addition, $N$ is planar (and $H(s) \not\equiv 0$), then $N$ has a \emph{dual} network $N^{d}$ whose impedance is $1/H(s)$. Hence, any given network class $\mathcal{N}$ induces a second (possibly identical) class $\mathcal{N}^{i}$, containing the frequency inverted network of every single network from $\mathcal{N}$. If, in addition, the networks in $\mathcal{N}$ are planar, then $\mathcal{N}$ induces two further classes: (i) $\mathcal{N}^{d}$, containing the dual network of every single network from $\mathcal{N}$; and (ii) $\mathcal{N}^{di} \coloneqq (\mathcal{N}^{d})^{i}$. Thus, $\mathcal{N}$ induces the network \emph{quartet} $\mathcal{Q}$, which is the union of $\mathcal{N}, \mathcal{N}^{i}, \mathcal{N}^{d}$, and $\mathcal{N}^{di}$ if the networks in $\mathcal{N}$ are planar; and the union of $\mathcal{N}$ and $\mathcal{N}^{i}$ otherwise.

\section{Hierarchical analysis of RLC Networks}
\label{sec:tnab}
In \cite{HugNa, HugTh}, a framework for the analysis of RLC networks is presented. This framework is influenced by the behavioral approach to dynamical systems \cite{JWBAOIS}, and graph theory results from \cite{BB_MGT} (we note also reference \cite{VdSphsog}, which applies graph theory results from \cite{BB_MGT} to the analysis of port-Hamiltonian systems). In this Appendix, we summarise relevant results from \cite[Part 1]{HugTh}, and we refer to \cite{HugTh} for detailed proofs.
\begin{remunerate}
\labitem{\ref{sec:tnab}\arabic{muni}}{nl:gt0-} A graph is a pair $(V, E)$ where $V$ is a set $\lbrace x_{1}, \ldots , x_{n}\rbrace$ whose elements are called \emph{vertices} and $E$ is a set of unordered pairs of vertices called edges, i.e., $E = \lbrace y_{1}, \ldots , y_{q}\rbrace$ where $y_{k} = (x_{k_{1}}, x_{k_{2}})$ for $k = 1, \ldots , q$ and for some $k_{1}, k_{2} \in 1, \ldots n$. A sequence of edges in a graph between two vertices $x_{a}$ and $x_{b}$ is called a path from $x_{a}$ to $x_{b}$. A \emph{circuit} is a path from a vertex $x_{a}$ to itself in which all of the edges are distinct and no vertices other than $x_{a}$ are repeated. A graph is called \emph{connected} if, for any given pair of vertices $x_{a}$ and $x_{b}$, there is a path from $x_{a}$ to $x_{b}$. A \emph{cut} in a connected graph is a set of edges whose removal partitions the vertices into two disjoint sets. It is called a \emph{cut-set} if, in addition, it contains no subset which is also a cut. A graph is called oriented when each edge has one of its two vertices assigned as a head vertex and the other as a tail vertex (we say the edge is oriented towards the head vertex). 

\labitem{\ref{sec:tnab}\arabic{muni}}{nl:gt0} Following \cite[Section 3]{HugNa}, we associate any given RLC network $N$ with a connected oriented graph $G$ which has two designated driving-point vertices and contains edges $y_{1}, \ldots , y_{m}$ corresponding to the elements $\hat{N}_{1},  \ldots , \hat{N}_{m}$ in the network. The edge $y_{k}$ has a current $i_{k}$, voltage $v_{k}$, and a relationship $p_{k}(\tfrac{d}{dt})i_{k} = q_{k}(\tfrac{d}{dt})v_{k}$ corresponding to the properties of element $\hat{N}_{k}$ ($k = 1, \ldots , m$). There is one additional edge $y_{0}$ in $G$ which is incident with the two driving-point vertices, and has a current $-i$ and voltage $v$. This corresponds to a source being connected to $N$ as in Fig.\ \ref{fig:open} (see Section \ref{sec:ssrrlc}).

\labitem{\ref{sec:tnab}\arabic{muni}}{nl:gt0+} For the RLC network $N$ in \ref{nl:gt0}, we let
\begin{equation}
\hspace*{-0.25cm} \mathbf{i} \coloneqq \text{col}\!\begin{pmatrix}-i&\hspace{-0.15cm}  i_{1}&\hspace{-0.15cm} \cdots &\hspace{-0.15cm} i_{m}\end{pmatrix}, \hspace{0.1cm} \text{and } \mathbf{v} \coloneqq \text{col}\!\begin{pmatrix}v&\hspace{-0.15cm} v_{1}&\hspace{-0.15cm} \cdots &\hspace{-0.15cm} v_{m}\end{pmatrix}.\label{eq:ivvd}
\end{equation}
If (i) $p_{k}\left(\frac{d}{dt}\right)i_{k} = q_{k}\left(\frac{d}{dt}\right)v_{k}$ ($k = 1, \ldots , m$), and (ii) $\mathbf{i}(t)$ satisfies Kirchhoff's current law and $\mathbf{v}(t)$ satisfies Kirchhoff's voltage law for all $t \in \mathbb{R}$, then we call $\text{col}\!\begin{pmatrix}i&\hspace{-0.15cm} v &\hspace{-0.15cm}  i_{1}&\hspace{-0.15cm} \cdots &\hspace{-0.15cm} i_{m}&\hspace{-0.15cm} v_{1}&\hspace{-0.15cm} \cdots &\hspace{-0.15cm} v_{m}\end{pmatrix}$ a \emph{trajectory} of $N$, with $\text{col}\!\begin{pmatrix}i&\hspace{-0.15cm} v\end{pmatrix}$ the corresponding \emph{driving-point trajectory}. The \emph{behavior} (resp., \emph{driving-point behavior}) of $N$ is defined as the set of all network \emph{trajectories} (resp., \emph{driving-point trajectories}). 

\labitem{\ref{sec:tnab}\arabic{muni}}{nl:gt1-} Kirchhoff's laws are related to the cut-set and circuit spaces of a graph (see \cite{HugNa}). These two spaces correspond to fundamental subspaces of the graph's incidence matrix. Specifically, let $N$ and $G$ be as in \ref{nl:gt0}. The incidence matrix $M$ of $G$ is an $n \times (m+1)$ matrix whose $ij$th entry is $-1$ (resp., $+1$) if edge $y_{j-1}$ is incident with vertex $x_{i}$ and oriented towards (resp., away from) $x_{i}$, and $0$ otherwise. It can then be shown that the $\mathbb{R}$-vector space spanned by the rows of $M$ is the \emph{cut-set space} of $G$ (and has dimension $n-1$ as $G$ is connected); the orthogonal $\mathbb{R}$-vector space $\lbrace \mathbf{z} \in \mathbb{R}^{m+1} \mid M\mathbf{z} = 0\rbrace$ is the \emph{circuit space} of $G$; and Kirchhoff's current (resp., voltage) law implies that $\mathbf{i}(t)$ (resp., $\mathbf{v}(t)$) is in the circuit (resp., cut-set) space of $G$ for all $t \in \mathbb{R}$ \cite{HugNa, HugTh}.

\labitem{\ref{sec:tnab}\arabic{muni}}{nl:gt1} A vertex in a connected graph whose deletion renders the graph disconnected is called a \emph{cut vertex}. A graph (or an RLC network) is called \emph{biconnected} if it is connected and it has no cut vertices. A \emph{biconnected component} of a graph $G$ is a biconnected subgraph of $G$ which is not a subgraph of any larger biconnected subgraph of $G$. Note, if $G$ contains an edge which is incident with a single vertex (a loop), then this edge is a biconnected component. It can be shown that the driving-point behavior of an RLC network $N$ is unchanged by removing elements which are not in the biconnected component of $N$ containing the source \cite[proof of Lemma 1.2.3]{HugTh}. Consequently, we restrict our attention in this paper to biconnected networks.

\labitem{\ref{sec:tnab}\arabic{muni}}{nl:gt2} We consider two networks to be identical if there is an ordering and orientation of the edges in their respective graphs such that (i) the two graphs have the same circuit space and cut-set space; and (ii) the relationships associated with the respective edges are identical. With this ordering and orientation, the two networks have the same behavior.

\labitem{\ref{sec:tnab}\arabic{muni}}{nl:gt3-} Let $N$ be an RLC network and $G$ the corresponding graph (as in \ref{nl:gt0}). A non-empty subset $\hat{n}$ of the elements in $N$ is called a \emph{subnetwork} if the corresponding edges form a connected subgraph of $G$ (we do not allow a subnetwork to contain the source). Let $\hat{N}$ be a subnetwork of $N$ which contains exactly two vertices $x_{a}$ and $x_{b}$ where the source and/or elements in $N$ but not in $\hat{N}$ are incident. Then $\hat{N}$ is an RLC network with driving-point terminals $x_{a}$ and $x_{b}$, and we call $\hat{N}$ a \emph{one-port} (subnetwork) in $N$. For any given trajectory of $N$, we define the current $\hat{i}$ and voltage $\hat{v}$ in $\hat{N}$ as follows. We let $\hat{i} \coloneqq \hat{i}_{+} - \hat{i}_{-}$, where $\hat{i}_{+}$ (resp., $\hat{i}_{-}$) is the sum of the currents through the elements in $\hat{N}$ which are incident with $x_{a}$ and oriented away from (resp., towards) $x_{a}$. To define $\hat{v}$, we pick an arbitrary path in $\hat{N}$ from $x_{a}$ to $x_{b}$, and we let $\hat{v} \coloneqq \hat{v}_{+} - \hat{v}_{-}$, where $\hat{v}_{+}$ (resp., $\hat{v}_{-}$) is the sum of the voltages across the elements in the path which are oriented with (resp., against) the path. It follows from \cite[proof of Theorem 1.9.6]{HugTh} that $\hat{v}$ does not depend on the choice of path, and $\text{col}\!\begin{pmatrix}\hat{i}& \hspace{-0.15cm} \hat{v}\end{pmatrix}$ is a driving-point trajectory for $\hat{N}$.

\labitem{\ref{sec:tnab}\arabic{muni}}{nl:gt3} We say an RLC network $N$ \emph{comprises} the one-ports $\hat{N}_{1}, \ldots , \hat{N}_{m}$ if each element in $N$ belongs to one and only one of these $m$ one-ports. There is an associated graph $\tilde{G}$ which is obtained from the graph $G$ described in \ref{nl:gt0} by replacing the edges in $G$ corresponding to the elements in $\hat{N}_{k}$ by a single edge $y_{k}$ between the driving-point terminals of $\hat{N}_{k}$ ($k = 1, \ldots , m$). Since $N$ is biconnected (see \ref{nl:gt1}), it is easily shown that $\tilde{G}$ is too. Now, consider a trajectory of $N$, let $i_{k}$ denote the current and $v_{k}$ the voltage in $\hat{N}_{k}$ ($k = 1, \ldots , m$), let $-i$ denote the current and $v$ the voltage in the source, and let $\mathbf{i}$ and $\mathbf{v}$ be as in (\ref{eq:ivvd}). Then it can be shown that $\mathbf{i}(t)$ (resp., $\mathbf{v}(t)$) is in the circuit (resp., cut-set) space of $\tilde{G}$ for all $t \in \mathbb{R}$ \cite[proof of Theorem 1.9.6]{HugTh}. Also, if (i) $\mathbf{i}(t)$ is in the circuit space and $\mathbf{v}(t)$ is in the cut-set space of $\tilde{G}$ for all $t \in \mathbb{R}$, and (ii) $\text{col}\!\begin{pmatrix}i_{k}& \hspace{-0.15cm} v_{k}\end{pmatrix}$ is a driving-point trajectory of $\hat{N}_{k}$ ($k = 1, \ldots , m$), then $\text{col}\!\begin{pmatrix}i& \hspace{-0.15cm} v\end{pmatrix}$ is a driving-point trajectory of $N$. In particular, since the cut-set and circuit spaces of a graph are orthogonal (see \ref{nl:gt1-}), then $\mathbf{i}^{T}(t)\mathbf{v}(t) = 0$ for all $t \in \mathbb{R}$.

\labitem{\ref{sec:tnab}\arabic{muni}}{nl:gt3b} From \ref{nl:gt3}, it is easily shown that the driving-point behavior (resp., impedance) of an RLC network $N$ is unchanged if we replace a one-port $\hat{N}$ in $N$ with a network which has the same driving-point behavior (resp., impedance) as $\hat{N}$.

\labitem{\ref{sec:tnab}\arabic{muni}}{nl:gt6} Let $N$ be an RLC network (following \ref{nl:gt1}, $N$ is biconnected). We say that $N$ is a series (resp., parallel) connection of two RLC networks $\hat{N}_{1}$ and $\hat{N}_{2}$ if (i) $\hat{N}_{1}$ and $\hat{N}_{2}$ are both one-ports in $N$; (ii) all the elements in $N$ are either in $\hat{N}_{1}$ or $\hat{N}_{2}$; and (iii) there is exactly one vertex (resp., two vertices) where elements from both $\hat{N}_{1}$ and $\hat{N}_{2}$ are incident. It is then easily shown from \ref{nl:gt3b} that $H(s) = Z_{1}(s) + Z_{2}(s)$ (resp., $1/H(s) = 1/Z_{1}(s) + 1/Z_{2}(s)$) , where $H(s), Z_{1}(s)$, and $Z_{2}(s)$ denote the impedances of $N, \hat{N}_{1}$, and $\hat{N}_{2}$, respectively.

\labitem{\ref{sec:tnab}\arabic{muni}}{nl:gt8} A trajectory in which the currents and voltages in the network are all varying sinuoidally at a fixed but arbitrary frequency $\omega \in \mathbb{R}$ is called a \emph{sinusoidal trajectory} (at frequency $\omega$). The corresponding driving-point trajectory is called a \emph{sinusoidal driving-point trajectory}. The existence of a non-zero sinusoidal driving-point trajectory at frequency $\omega$ for any given RLC network $N$ and $\omega \in \mathbb{R}$ is guaranteed by \cite[Theorem 5]{HugNa}. Consider a sinusoidal trajectory of $N$ and a one-port $\hat{N}_{k}$ in $N$. It is easily shown that the current and voltage in $\hat{N}_{k}$ are also varying sinusoidally (and correspond to a sinusoidal driving-point trajectory of $\hat{N}_{k}$). In other words, there exist $\tilde{i}_{k}, \tilde{v}_{k} \in \mathbb{C}$ such that the current $i_{k}$ and voltage $v_{k}$ in $\hat{N}_{k}$ satisfy $i_{k}(t) = \Re{(\tilde{i}_{k}e^{j\omega t})}$ and $v_{k}(t) = \Re{(\tilde{v}_{k}e^{j\omega t})}$ for all $t \in \mathbb{R}$. We call $\tilde{i}_{k}$ the phasor current, and $\tilde{v}_{k}$ the phasor voltage, of $\hat{N}_{k}$ (corresponding to this specific sinusoidal trajectory). Also, the driving-point current $i$ and voltage $v$ take the forms $i(t) = \Re{(\tilde{i}e^{j\omega t})}$ and $v(t) = \Re{(\tilde{v}e^{j\omega t})}$, respectively, for all $t \in \mathbb{R}$ and for some $\tilde{i}, \tilde{v} \in \mathbb{C}$, and we call $\tilde{i}$ the phasor current,  and $\tilde{v}$ the phasor voltage, of the source. Finally, denoting the impedance of the network by $H(s)$, then $\tilde{i} = 0$ if $H(s)$ has a pole at $s = j\omega$, with $\tilde{v} = H(j\omega)\tilde{i}$ otherwise \cite[Theorem 5]{HugNa}.

\labitem{\ref{sec:tnab}\arabic{muni}}{nl:gt7} Let $N$ be an RLC network, and let $\hat{N}$ be a one-port in $N$ with driving-point vertices $x_{a}$ and $x_{b}$. By \emph{opening} (resp., \emph{shorting}) $\hat{N}$ in $N$, we mean the operation of removing all of the elements in $\hat{N}$ from $N$ (resp., connecting the two vertices $x_{a}$ and $x_{b}$ in $N$), and then removing all elements which are not in the same biconnected component as the source. We note that the resulting network $N_{a}$ could contain no elements, with the source incident with two distinct vertices (resp., two coincident vertices), in which case $N_{a}$ represents an \emph{open circuit} (resp., \emph{short circuit}), and $N_{a}$ does not possess an impedance (resp., the impedance of $N_{a}$ is identically zero). 

\labitem{\ref{sec:tnab}\arabic{muni}}{nl:gt7+} Let $N$ be an RLC network comprising the elements $\hat{N}_{1}, \ldots , \hat{N}_{m}$; let $\tilde{N}$ be a one-port in $N$; let $N_{a}$ be obtained by opening (resp., shorting) the one-port $\tilde{N}$ in $N$, and (without loss of generality) let $N_{a}$ comprise the elements $\hat{N}_{1}, \ldots \hat{N}_{r}$; and let $\mathbf{b} \coloneqq \text{col}\!\begin{pmatrix}i&\hspace{-0.15cm} v &\hspace{-0.15cm}  i_{1}&\hspace{-0.15cm} \cdots &\hspace{-0.15cm} i_{m}&\hspace{-0.15cm} v_{1}&\hspace{-0.15cm} \cdots &\hspace{-0.15cm} v_{m}\end{pmatrix}$ be a sinusoidal trajectory of $N$ at frequency $\omega$. If the phasor current (resp., voltage) in $\tilde{N}$ is zero, then $\mathbf{b}_{a} := \text{col}\!\begin{pmatrix}i&\hspace{-0.15cm} v &\hspace{-0.15cm}  i_{1}&\hspace{-0.15cm} \cdots &\hspace{-0.15cm} i_{r}&\hspace{-0.15cm} v_{1}&\hspace{-0.15cm} \cdots &\hspace{-0.15cm} v_{r}\end{pmatrix}$ is a sinusoidal trajectory of $N_{a}$ \cite[proof of Lemma 3.5.10]{HugTh}. In particular, if $\bar{N}$ is a one-port in $N$ and also in $N_{a}$, then it has the same phasor current and voltage in the two sinusoidal trajectories $\mathbf{b}$ and $\mathbf{b}_{a}$, so we may repeat this process with the one-port $\bar{N}$ if its phasor current or voltage is also zero.

\labitem{\ref{sec:tnab}\arabic{muni}}{nl:gt9}  Let $N$ be an RLC network with impedance $H(s)$. We say that $N$ has a driving-point C-cut-set (resp., L-cut-set) if removal of all the capacitors (resp., inductors) in the network leaves the driving-point terminals disconnected, and a driving-point C-path (resp., L-path) if there is a path between the driving-point terminals comprised solely of capacitors (resp., inductors). It is well known that (i) $H(s)$ has a pole at $s=0$ (resp., $s = \infty$) if $N$ has a driving-point C-cut-set (resp., L-cut-set); and (ii) $H(\infty) = 0$ (resp., $H(0) = 0$) if $N$ has a driving-point C-path (resp., L-path) \cite[Theorem 8.3]{Seshu_Reed}. Now, suppose $H(s)$ does not have a pole at $s = 0$ (resp., $s = \infty$), and let $N_{a}$ be the network obtained by either opening (resp., shorting) a capacitor in $N$, or shorting (resp., opening) an inductor in $N$. Then $N_{a}$ has impedance $H_{a}(s)$ which satisfies $H_{a}(0) = H(0)$ (resp., $H_{a}(\infty) = H(\infty)$) \cite[Corollary 3.5.11]{HugTh}.
\end{remunerate}

\section{Synthesis of passive mechanical controllers}
\label{sec:spmc}
Figure \ref{fig:ema} (see Section \ref{sec:ssrrlc}) indicates the properties of the two-terminal mechanical components: dampers, springs, and inerters \cite{mcs02}. Using the force-current analogy, there is a one to one correspondence between these elements and the electrical elements resistors, inductors, and capacitors (see Fig.\ \ref{fig:ema}). This analogy extends to the interconnection laws: the net sum of all currents/forces at any vertex is zero; and the net sum of all voltages/velocities around any circuit is zero. 

Note that there are restrictions to the analogy between a damper-spring-mass network and an RLC network \cite{mcs02}. Specifically, the force applied to an inerter is proportional to the relative acceleration of its two terminals, whereas for the mass it is proportional to its acceleration \emph{relative to ground} (formally, the fixed point in the inertial reference frame). Consequently, a mass is analogous to a grounded capacitor, so damper-spring-mass networks are analogous to RLC networks in which \emph{all capacitors are grounded}. On the other hand, every single RLC network has an equivalent damper-spring-inerter network. The transfer function from the force applied to the damper-spring-inerter network to the relative velocity of the network's terminals is equivalent to the impedance of the corresponding RLC network.
 
For reasons of cost, complexity, reliability, regulations, and power requirements, it is often desirable in mechanical applications to use a passive controller such as a damper-spring-inerter network. Applications of damper-spring-inerter networks to vehicle suspension, train suspension, motorcycle steering compensators, and building suspension are described in \cite{mcs02, chen_14, fucheng_10, fucheng_12, jiang_vsd_12, Limebeer_steering2, Limebeer_Steering, fucheng_07}. The present paper considered the realization of a PR impedance using the minimum possible number of elements, and is therefore relevant to the design of passive mechanical controllers.

\bibliographystyle{IEEEtran}
\bibliography{IEEEabrv,wmrte_thh}

\newcommand{\noopsort}[1]{} \newcommand{\singleletter}[1]{#1}
\begin{thebibliography}{10}
\providecommand{\url}[1]{#1}
\csname url@samestyle\endcsname
\providecommand{\newblock}{\relax}
\providecommand{\bibinfo}[2]{#2}
\providecommand{\BIBentrySTDinterwordspacing}{\spaceskip=0pt\relax}
\providecommand{\BIBentryALTinterwordstretchfactor}{4}
\providecommand{\BIBentryALTinterwordspacing}{\spaceskip=\fontdimen2\font plus
\BIBentryALTinterwordstretchfactor\fontdimen3\font minus
  \fontdimen4\font\relax}
\providecommand{\BIBforeignlanguage}[2]{{%
\expandafter\ifx\csname l@#1\endcsname\relax
\typeout{** WARNING: IEEEtran.bst: No hyphenation pattern has been}%
\typeout{** loaded for the language `#1'. Using the pattern for}%
\typeout{** the default language instead.}%
\else
\language=\csname l@#1\endcsname
\fi
#2}}
\providecommand{\BIBdecl}{\relax}
\BIBdecl

\bibitem{JWBAOIS}
J.~C. Willems, ``The behavioral approach to open and interconnected systems,''
  \emph{Control Systems Magazine}, vol.~27, pp. 46--99, 2007.

\bibitem{Foster_24}
R.~M. Foster, ``A reactance theorem,'' \emph{Bell Syst. Techn. Journ.}, vol.~3,
  p. 259, 1924.

\bibitem{Brune}
O.~Brune, ``Synthesis of a finite two-terminal network whose driving-point
  impedance is a prescribed function of frequency,'' \emph{J. Math. Phys.},
  vol.~10, pp. 191--236, 1931.

\bibitem{kalman2010}
R.~E. Kalman, ``Old and new directions of research in system theory,''
  \emph{Perspectives in Mathematical System Theory, Control, and Signal
  Processing}, vol. 398, pp. 3--13, 2010.

\bibitem{HugJSmOp}
\BIBentryALTinterwordspacing
T.~H. Hughes, J.~Z. Jiang, and M.~C. Smith. (2014) Two problems on minimality
  in {R}{L}{C} circuit synthesis, workshop on `{D}ynamics and {C}ontrol in
  {N}etworks', {L}und {U}niversity. [Online]. Available:
  \url{http://www.lccc.lth.se/media/2014/malcolm.pdf}
\BIBentrySTDinterwordspacing

\bibitem{camwb}
M.~K. \c{C}amlibel, J.~C. Willems, and M.~N. Belur, ``On the dissipativity of
  uncontrollable systems,'' in \emph{Proceedings of the 42nd IEEE Conference on
  Decision and Control, Hawaii}, Dec. 2003.

\bibitem{JWDDS}
J.~C. Willems, ``Dissipative dynamical systems,'' \emph{European Journal on
  Control}, vol.~13, pp. 134--151, 2007.

\bibitem{JW_HVDS}
------, ``Hidden variables in dissipative systems,'' \emph{Proceedings of the
  43rd IEEE Conference on Decision and Control}, pp. 358--363, 2004.

\bibitem{BD}
R.~Bott and R.~J. Duffin, ``Impedance synthesis without use of transformers,''
  \emph{J. Appl. Phys.}, vol.~20, p. 816, 1949.

\bibitem{Reza_54}
F.~M. Reza, ``Synthesis without ideal transformers,'' \emph{J. Appl. Phys.},
  vol.~25, pp. 807--808, March 1954.

\bibitem{PS}
R.~H. Pantell, ``A new method of driving point impedance synthesis,''
  \emph{Proc.\ IRE (Correspondence)}, vol.~42, p. 861, 1954.

\bibitem{Fialkow_Gerst}
A.~Fialkow and I.~Gerst, ``Impedance synthesis without mutual coupling,''
  \emph{Quart. Appl. Math.}, vol.~12, pp. 420--422, 1955.

\bibitem{HugSmSP}
T.~H. Hughes and M.~C. Smith, ``On the minimality and uniqueness of the
  {B}ott-{D}uffin realization procedure,'' \emph{IEEE Trans. on Automatic
  Control}, vol.~59, no.~7, pp. 1858--1873, July 2014.

\bibitem{mcs02}
M.~C. Smith, ``Synthesis of mechanical networks: the inerter,'' \emph{IEEE
  Trans. on Automatic Control}, vol.~47, no.~10, pp. 1648--1662, 2002.

\bibitem{chen_14}
Y.~Hu, M.~Z.~Q. Chen, and Z.~Shu, ``Passive vehicle suspensions employing
  inerters with multiple performance requirements,'' \emph{Journal of Sound and
  Vibration}, vol. 333, no.~8, pp. 2212--2225, 2014.

\bibitem{fucheng_10}
F.~C. Wang and M.~K. Liao, ``The lateral stability of train suspension systems
  employing inerters,'' \emph{Vehicle System Dynamics}, vol.~48, no.~5, pp.
  619--643, 2010.

\bibitem{fucheng_12}
F.~C. Wang, M.~R. Hsieh, and H.~J. Chen, ``Stability and performance analysis
  of a full-train system with inerters,'' \emph{Vehicle System Dynamics},
  vol.~50, no.~4, pp. 545--571, 2012.

\bibitem{jiang_vsd_12}
J.~Z. Jiang, A.~Z. Matamoros-Sanchez, R.~M. Goodall, and M.~C. Smith, ``Passive
  suspensions incorporating inerters for railway vehicles,'' \emph{Vehicle
  System Dynamics. Special Issue: IAVSD Supplement}, vol.~50, pp. 263--276,
  2012.

\bibitem{Limebeer_steering2}
S.~Evangelou, D.~J.~N. Limebeer, R.~S. Sharp, and M.~C. Smith, ``Control of
  motorcycle steering instabilities - passive mechanical compensators
  incorporating inerters,'' \emph{IEEE Control Systems Magazine}, pp. 78--88,
  October 2006.

\bibitem{Limebeer_Steering}
------, ``Mechanical steering compensation for high-performance motorcycles,''
  \emph{Transactions of ASME, J. of Applied Mechanics}, vol.~74, no.~2, pp.
  332--346, 2007.

\bibitem{fucheng_07}
F.~C. Wang, M.~F. Hong, and C.~W. Chen, ``Performance analyses of building
  suspension control with inerters.''\hskip 1em plus 0.5em minus 0.4em\relax
  IEEE conference on Decision and Control, Dec. 2007, pp. 3786--3791.

\bibitem{JiangSmith11}
J.~Z. Jiang and M.~C. Smith, ``Regular positive-real functions and five-element
  network synthesis for electrical and mechanical networks,'' \emph{IEEE Trans.
  on Automatic Control}, vol.~56, no.~6, pp. 1275--1290, June 2011.

\bibitem{BelOc}
R.~U. Chavan, V.~P. Samuel, K.~Mallick, and M.~N. Belur, ``Optimal
  charging/discharging and commutativity properties of {A}{R}{E} solutions for
  {R}{L}{C} circuits,'' \emph{Proc.\ of the 21th International Symposium on
  Mathematical Theory of Networks and Systems, Groningen, Netherlands}, July
  2014.

\bibitem{HugNa}
T.~H. Hughes and M.~C. Smith, ``Controllability of linear passive network
  behaviors,'' \emph{In press, Systems and Control Letters,
  doi:10.1016/j.sysconle.2015.09.011}, 2015.

\bibitem{HugSmAI}
------, ``Algebraic criteria for circuit realisations,'' in \emph{Mathematical
  System Theory - Festschrift in Honor of Uwe Helmke on the Occasion of his
  Sixtieth Birthday}, K.~Huper and J.~Trumpf, Eds.\hskip 1em plus 0.5em minus
  0.4em\relax CreateSpace, 2012.

\bibitem{Smf}
S.~Seshu, ``Minimal realizations of the biquadratic minimum function,''
  \emph{IRE Trans. Circuit Theory}, vol.~6, no.~4, pp. 345--350, 1959.

\bibitem{HugTh}
T.~H. Hughes, ``On the synthesis of passive networks without transformers,''
  Ph.D. dissertation, {U}niversity of {C}ambridge, 2014.

\bibitem{Storer_54}
J.~E. Storer, ``Relationship between the {B}ott-{D}uffin and {P}antell
  impedance synthesis,'' \emph{Proc.\ IRE}, vol.~42, p. 1451, September 1954.

\bibitem{Foster_63c}
R.~M. Foster, ``Minimum biquadratic impedances,'' \emph{IEEE Trans. on Circuit
  Theory}, vol.~10, no.~4, p. 527, 1963.

\bibitem{Telge}
B.~D.~H. Tellegen, ``Geometrical configurations and duality of electrical
  networks,'' \emph{Philips Technical Review}, vol.~5, pp. 324--330, 1940.

\bibitem{HugCF}
T.~H. Hughes, ``On connections between the {C}auchy index, the {S}ylvester
  matrix, continued fraction expansions, and circuit synthesis,'' \emph{Proc.\
  of the 21th International Symposium on Mathematical Theory of Networks and
  Systems, Groningen, Netherlands}, July 2014.

\bibitem{BB_MGT}
B.~Bollob\'{a}s, \emph{Modern graph theory}.\hskip 1em plus 0.5em minus
  0.4em\relax New York : Springer, 1998.

\bibitem{VdSphsog}
A.~J. van~der Schaft and B.~M. Maschke, ``Port-{H}amiltonian systems on
  graphs,'' \emph{SIAM Journal on Control Optim.}, vol.~51, no.~2, pp.
  906--937, 2013.

\bibitem{Seshu_Reed}
S.~Seshu and M.~B. Reed, \emph{Linear Graphs and Electrical Networks}.\hskip
  1em plus 0.5em minus 0.4em\relax Addison-Wesley, 1961.

\end{thebibliography}

\begin{IEEEbiography}[{\includegraphics[width=1in,height=1.25in,clip,keepaspectratio]{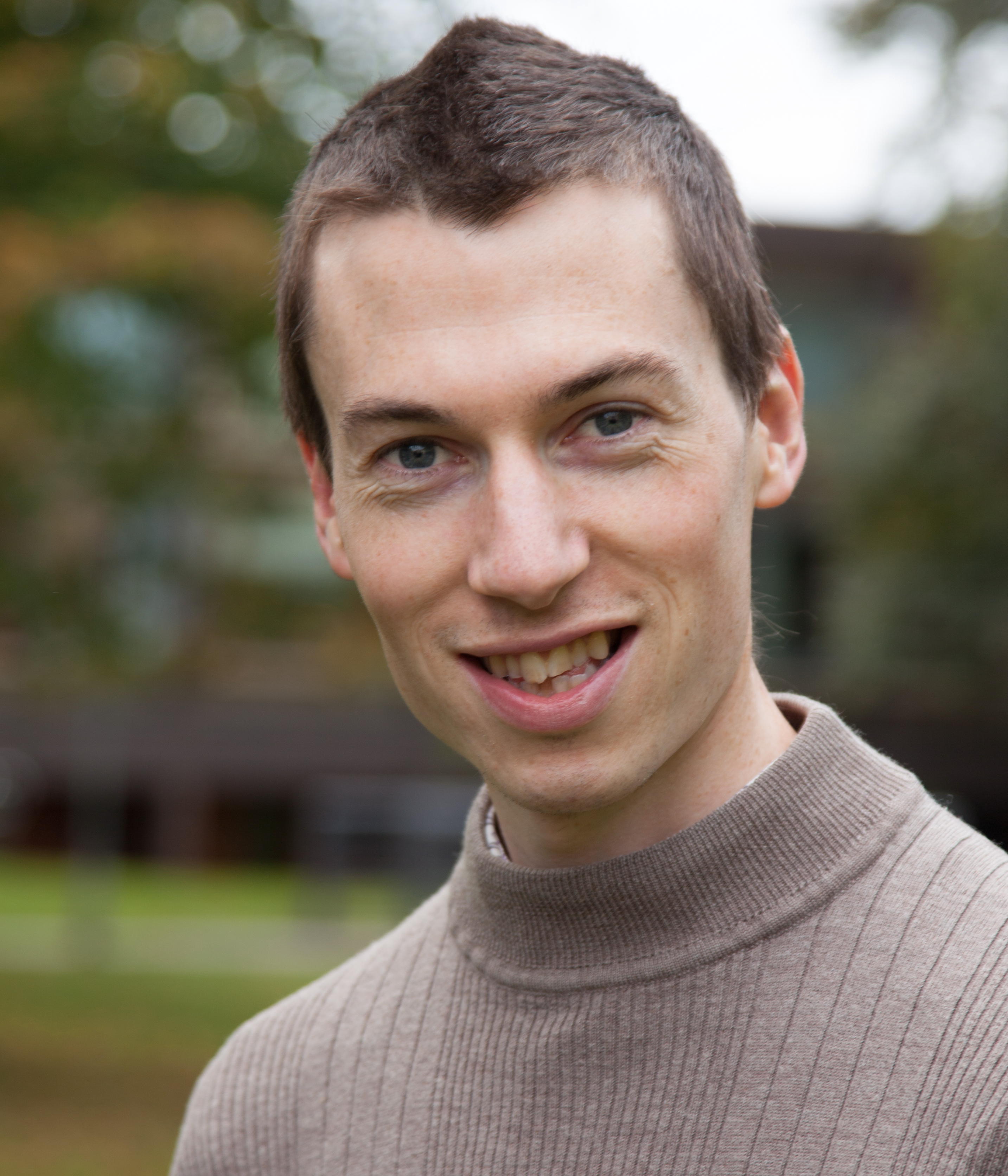}}]{Timothy H. Hughes}
received the M.Eng. degree in mechanical engineering, and the Ph.D degree in control engineering, from the University of Cambridge, U.K., in 2007 and 2014, respectively. From 2007 to 2010 he was employed as a mechanical engineer at The Technology Partnership, Hertfordshire, U.K. He is currently the Henslow Research Fellow at Fitzwilliam College, University of Cambridge.

He has a general interest in systems and control theory, and a specific interest in passive mechanical and electrical control and network synthesis.
\end{IEEEbiography}
\vfill

\end{document}